\documentclass[journal,10pt]{IEEEtran}
\usepackage{amsmath,amssymb,latexsym,mathrsfs,bm,cite,color,amsthm}
\usepackage{graphicx}
\usepackage{multirow}
\usepackage{array}
\usepackage{mdwmath}
\usepackage{mdwtab}
\usepackage{algorithm}
\usepackage{algorithmic}
\usepackage[tight,footnotesize]{subfigure}
\usepackage{lipsum} 
\usepackage{stfloats}
\usepackage{tabularx}
\usepackage{booktabs}
\newtheorem{theorem}{Theorem}
\newtheorem{lemma}{Lemma}

\begin{document}

\title{Wireless Channel Delay Spread Performance Evaluation of a Building Layout}

\author{Yixin Huang,~
        Jiliang Zhang,~\textit{Senior Member, IEEE},
        and~Jie Zhang,~\textit{Senior Member, IEEE}
        \vspace{-20pt}
\thanks{Yixin Huang is with the Department of Electronic and Electrical Engineering, The University of Sheffield, Sheffield S1 4ET, UK. 

Jiliang Zhang and Jie Zhang are with the Department of Electronic and Electrical Engineering, the University of Sheffield, Sheffield, S1 4ET, UK, and also with Ranplan Wireless Network Design Ltd., Cambridge, CB23 3UY, UK (email: jiliang.zhang@sheffield.ac.uk).}
}

\markboth{IEEE}
{}

\maketitle

\begin{abstract}

Evaluating a building's wireless performance during the building design process is {\color{black}  a new paradigm for wireless communications and building design.} This paper proposes the earliest building wireless performance (BWP) evaluation theory focusing on the channel delay spread (DS). The novel contributions of this paper lie in the following aspects. 1){\color{black} We define a new metric called DS gain, which is the first metric for evaluating the channel DS performance of a building under design.} 2) We propose an analytical model to compute the metric quickly and accurately. 3) The proposed scheme is validated via Monte-Carlo simulations under typical indoor scenarios. Numerical results reveal that building design has a clear impact on the root mean square (RMS) DS. In the future, architects need to design a building taking its DS gain into account carefully. Otherwise, indoor networks in it will suffer from severe signal inter-symbol interference (ISI) due to an over large DS gain.
\end{abstract}


\IEEEpeerreviewmaketitle

\section{Introduction}
\IEEEPARstart{T}{he} building wireless performance (BWP) has to be considered in the building design process \cite{zhang2020wireless,zhang2021fundamental,zhang2021wireless,yang2021partition}. The performance of the densely deployed indoor wireless devices is upper bounded by the building design regardless of specific network deployment and configuration. With the evolution of wireless reliability-dependent services such as augmented reality, virtual reality, smart building, smart factories, etc.\cite{3gpp22104,alliance20195g,3gpp22084,sabella2018industrial}, the wireless performance demand, e.g., throughput, latency, and reliability, increases without limitation and will eventually hit the upper bound. Therefore, expensive and time-consuming building modification may become the last resort to meet these wireless performance requirements in case the BWP is not considered in the building design process. {\color{black}As a result, BWP metrics, which can assist architects in avoiding employing room layouts with bad wireless performance, are essential in the building design stages.}

It is essential to consider delay spread (DS),  a key channel characteristics, in the BWP evaluation.  {\color{black}If a building is not well designed, considering its impact on channel DS, a strong inter-symbol interference (ISI) may occur \cite{rappaport1996wireless,ha2010theory,walrand2000high,diez2020reliability}, resulting in a narrower coherence bandwidth \cite{debaenst2020rms}. Since the transmitted signal will exhibit selective fading in the channel when the transmitted signal bandwidth is smaller than the coherence bandwidth, a bad building design may degrade the reliability of indoor wireless networks and reduce the upper-bound of indoor transmitted signal bandwidth \cite{3gpp38300,3gpp38913}. On the other hand, the broader the coherence bandwidth is, the more accurate the channel estimation in the wider frequency band and the easier the pilot design is. As a result, the building's inner layout will also affect the complexity and effectiveness of the pilot design \cite{shieh2009ofdm}. }

However, how to evaluate BWP considering channel DS is still unknown. Most existing studies on channel DS lie in the perspective of channel modeling rather than BWP evaluation \cite{hashemi1992analysis,hashemi1993impulse,hashemi1994statistical,varela2001rms,siamarou2001multipath,cassioli2013characterization,zahedi2016experimental,coko2017rms,sharma2018improved,yu2015antenna,sun2017path,yu2017measurement}. 
Whereas in the field of BWP, the impact of building layout on channel DS has never been considered before \cite{zhang2021fundamental,zhang2021wireless,yang2021partition}. 

From the perspective of channel modeling, the root-mean-square (RMS) DS, which is a key metric to evaluate the channel DS, has been measured, characterized, and modeled under typical indoor propagation conditions \cite{hashemi1992analysis,hashemi1993impulse,hashemi1994statistical,varela2001rms,siamarou2001multipath,cassioli2013characterization,zahedi2016experimental,coko2017rms,sharma2018improved,yu2015antenna,sun2017path,yu2017measurement}. 
In \cite{varela2001rms,siamarou2001multipath,cassioli2013characterization,zahedi2016experimental,coko2017rms,sharma2018improved}, the RMS-DS of indoor channels was measured and modeled in the ultra-high frequency (UHF) band, ultra-wideband (UWB) and visible light communication system. In \cite{varela2001rms,yu2015antenna,sun2017path,yu2017measurement}, RMS-DS was modeled in different room types such as office, corridor, and staircase. The results show that the channel modeling of RMS-DS in different environments often shows similar patterns.
Specifically, according to \cite{hashemi1994statistical,cassioli2013characterization,yu2015antenna,zahedi2016experimental,yu2017measurement}, RMS-DS in indoor scenarios has a stronger association with distance, or, more precisely, with path loss (PL) than they do in outdoor environments. Particularly, the results of \cite{yu2017measurement} demonstrate that indoor scenarios with a variety of room types and line-of-sight (LOS) or non-line-of-sight (NLOS) conditions can exhibit a wide variety of probability density distributions (PDFs) of the RMS-DS. Nevertheless, how to avoid strong DS when designing buildings has never been considered, which could make the building a bottleneck limiting the performance of 5G and 6G reliability-demand services.

From the perspective of BWP, the impact of a building on the interference and signal receiving power has been systematically investigated \cite{zhang2020wireless,zhang2021fundamental,zhang2021wireless,yang2021partition}.
In \cite{zhang2020wireless}, for the first time, we defined and modeled interference gain (IG) as an intrinsic metric that can be used to quantify BWP and, on this basis, proposed a new method to maximize it.
In \cite{zhang2021fundamental}, we proposed the BWP evaluation framework. In \cite{zhang2021wireless}, we defined the wireless performance indicators power gain (PG) that assess how the building structure affects the channel in terms of the received signal strength. In \cite{yang2021partition}, we proposed an analytical model of PG and IG for complex room structures and incorporated building materials and building structures into the evaluation model.
{\color{black}Inspired by our previous works, BWP evaluation has been researched in the fields of medium frequency broadcast and data-driven propagation \cite{liodakis2022effect,bakirtzis2022stochastic}.}
However, the published works on BWP have never taken channel DS into account.

To bridge the gap between the BWP evaluation and channel DS modeling, in this paper, we propose the earliest systematic BWP evaluation approach focusing on RMS-DS. More specifically, we define metric, namely DS gain, to capture the impact of a building on the RMS-DS of wireless connections in it. Then, we propose an analytical model of the RMS-DS to facilitate quick DS gain evaluation given a building layout. The contributions of this paper are three-fold:

\begin{itemize}
	\item The first evaluation approach for BWP in terms of channel DS is proposed. The associated evaluation metric, namely the DS gain, is defined to capture the impact of a building on the DS of indoor wireless transmission in it. 
	\item A statistical model for the proposed metric is also derived for quick and accurate metric calculation. The PDF of RMS-DS is deduced according to the room layout of the assessed building, and the analytic expression of the metric DS gain is derived.
	\item {\color{black}An analytical expression of the reliability metric is also derived for assessing the reliability of the proposed model. It can help assist architects in better using the proposed metric.}
	\item The proposed analytical model is verified by the Monte-Carlo simulation, and the results show that the analytic results match simulation results very well. 
\end{itemize}

The proposed approach provides architects a tool to quickly assess the RMS-DS performance of building layouts in the building design stage and the wireless system design stage. 

This paper is organized as follows. Section II introduces the definition of the metric of DS gain. Section III introduces the detailed derivation of analytic DS gain model. Section IV presents the Monte-Carlo simulation results and verifies the accuracy of the proposed model. Finally, Section V concludes the proposed evaluation approach.

\section{Definition of metric to evaluate building channel DS performance}

\begin{table}
	\caption{Notation definition}  
	\label{notation}
	\begin{tabular}{cl}
		\hline
		Notation& Description\\
		\hline
		$G_\tau$& DS gain\\
		$\tau$& RMS-DS\\
		$\tau_{\mathrm{I}}$& Instantaneous value of RMS-DS indoors\\
		$\tau_{\mathrm{O}}$& Instantaneous value of RMS-DS in the open space\\
		$d$& Distance between Tx and Rx\\
		$f_{\tau_\mathrm{I}}$& Indoor RMS-DS function of $d$\\
		$f_{\tau_\mathrm{O}}$& Open-space RMS-DS function of $d$\\
		$\mathcal{P}$& Probability of an event\\
		$p$& PDF of $d$\\
		$\iota$& The event that the TX is in the BUD\\
		$\chi$& The event that the Rx is in the BUD\\
		$\kappa$& The set of blockage situation\\
		$\upsilon$& The set of room types\\
		$g_{\mathbf{r}_\mathrm{T}}$& PDF of Tx locating at a position\\
		$r_\mathrm{A}$& Aspect ratio\\ 
		$\mathbf{r}$& The coordinates\\
		$N_\mathrm{r}$& The amount of rooms in the BUD\\
		$S$& The size of a room\\
		$i$& The room number of a room in the BUD\\
		$V$& The size of the BUD\\
		$\xi$& The event that the Tx is in a room\\
		$m$& The long edge of a room\\
		$l$& The short edge of a room\\
		$X$& The long edge of the BUD\\
		$Y$& The short edge of the BUD\\
		$\mathcal{C}$& The set of coordinates\\
		$L_{\kappa,\upsilon}$& Path loss of indoor transmission with definite\\&  transmission conditions\\
		$\mu$& Expectation\\
		$\sigma$& Standard deviation\\
		$\mu_{\kappa_i,\upsilon,\tau}$& Expectation value of RMS-DS indoors with definite\\ & transmission conditions\\
		$X$& Normally distributed random variables\\
		$G_{\tau,\mathrm{sim}}$& The simulation result of $G_\tau$\\
		$G_{\tau,\mathrm{ana}}$& The analytic result of $G_\tau$\\
		{\color{black}$\sigma$}& {\color{black}The reliability of the model}\\
		\hline
	\end{tabular}
\end{table}

We define the metric, the DS gain $G_\tau$, to evaluate the building wireless performance for RMS-DS:
\begin{equation}
\label{G_t}
G_\tau=\mathrm{E}[\tau_{\mathrm{I}}]-\mathrm{E}[\tau_{\mathrm{O}}],
\end{equation}
where $\tau_{\mathrm{I}}$ and $\tau_{\mathrm{O}}$ are instantaneous values of RMS-DS in indoor and in open space environments, and $\mathrm{E}[\tau_{\mathrm{I}}]$ denotes the expectation value of RMS-DS for the assessed indoor environment, and $\mathrm{E}[\tau_{\mathrm{O}}]$ denotes the expectation of RMS-DS for an open space with the same occupied area.
The indicator $G_\tau$ represents a building's inherent wireless performance characteristic.
It indicates the effective varying in RMS-DS levels when the network covers a building as opposed to open space.
{\color{black}Physically, more complicated building structures result in more severe multi-path effects, leading to greater RMS-DS and ISI, and finally results in narrower coherence bandwidth. }

According to measurement results \cite{varela2001rms,siamarou2001multipath,cassioli2013characterization,zahedi2016experimental,coko2017rms,sharma2018improved}, $\tau_{\mathrm{I}}$ is a random variable {\color{black}in any given indoor scenario}, and the distribution of RMS-DS for a specific link depends on the distance between the transmitter (Tx) and the receiver (Rx), denoted by $d$. {\color{black}Therefore,  $\mathrm{E}[\tau_{\mathrm{I}}]$ must be found with the given PDF of $\tau$ and $d$, as:
\begin{equation}
\begin{split}
\label{gderive}
\mathrm{E}[\tau_{\mathrm{I}}]= \int_0^{+\infty}\tau_{\mathrm{I}} \int_0^{+\infty}f_{\tau_{\mathrm{I}}}(d)p_{\iota,\chi}(d)\mathrm{d}d\mathrm{d}\tau_{\mathrm{I}},
\end{split}
\end{equation}
where $f_{\tau_\mathrm{I}}(d)$ denotes the indoor RMS-DS distribution function that varies with $d$, $\iota$ denotes the event that Tx is in the building under design (BUD), and $\chi$ denotes the event that Rx is in the BUD, and $p_{\iota,\chi}(d)$ denotes the PDF of the distance between Tx and Rx when they randomly locate in the BUD. }
As a contrast, {\color{black}given the same area compared with the indoor scenario,} $\tau_{\mathrm{O}}$ as the RMS-DS value in the open space, only depends on $d$, i.e., given a definite $d$, the value of $\tau_{\mathrm{O}}$ is determined. As the result, $\mathrm{E}[\tau_{\mathrm{O}}]$ can be found with the given PDF of $d$, {\color{black} as:
\begin{equation}
\begin{split}
\label{gderive}
\mathrm{E}[\tau_{\mathrm{O}}]=\int_0^{+\infty}  f_{\tau_\mathrm{O}}(d)p_{\iota,\chi}(d) \mathrm{d}d,
\end{split}
\end{equation}
where $f_{\tau_\mathrm{O}}(d)$ denotes the open-space RMS-DS function of $d$.}

Moreover, whether the transmission is LOS or NLOS impacts the distribution of RMS-DS substantially in an indoor environment. Measurement results also show that different room types, e.g., offices and corridors lead to different distributions of RMS-DS \cite{yu2017measurement}. Therefore, the distribution of $\tau_{\mathrm{I}}$ depends on  
i) the distribution of the transmission condition between Tx and Rx, in terms of the blockage effect and the room type,
ii) the distribution of Tx-Rx distance $d$, and
iii) the distribution of RMS-DS for a link given $d$ and definite transmission conditions.
Similarly, the distribution of $\tau_{\mathrm{O}}$ depends on  
i) the distribution of Tx-Rx distance $d$, and
ii) the RMS-DS in open space for a link given $d$.

To decouple the intrinsic property of the building and the specific deployment of indoor wireless devices, we need to assume that the Txs and the Rxs are randomly deployed in the building under consideration in the building design stage. Under this situation, $d$ is a random variable, too.

In a complex building environment, $G_\tau$ needs to be quantified in the building design stage, considering both the random position of transceivers and the blockage effect. 

\section{Analytical model of the DS gain}

To quickly and accurately compute the DS gain, in this section, the detailed procedure for deriving $G_\tau$ is provided, and the final analytic expression of $G_\tau$ is given. 
Table \ref{notation} describes the main notations that appear in this paper.

$\mathrm{E}[\tau_{\mathrm{I}}]$ and $\mathrm{E}[\tau_{\mathrm{O}}]$ are derived in the following subsections III-A and III-B.

\subsection{Derivation of $\mathrm{E}[\tau_{\mathrm{I}}]$}

Within a BUD,$f_{\tau_{\mathrm{I}}}(d)$ can be calculated as:
{\color{black}\begin{eqnarray}
\label{eq5p5}
f_{\tau_{\mathrm{I}}}(d)
=\int_{\mathbf{r}}\sum_{\kappa}\sum_{\upsilon}f_{\tau_\mathrm{I},\chi|d,\kappa,\upsilon,\mathbf{r}_{\mathrm{T}}}(\tau_{\mathrm{I}})\mathcal{P}(\upsilon)\mathcal{P}(\kappa|d)
g_{\mathbf{r}_{\mathrm{T}}}(\mathbf{r})\mathrm{d}\mathbf{r},
\end{eqnarray}
where $\mathcal{P}(\cdot)$ denotes the probability of an event, $\kappa\in\{\mathrm{LOS},\mathrm{NLOS}\}$ denotes the blockage effect, $\upsilon\in\{\mathrm{office},\mathrm{hall},\mathrm{corridor},\mathrm{staircase},...\}$ denots the type of the room where the Tx locates, $\mathbf{r}_{\mathrm{T}}=(x_{\mathrm{T}},y_{\mathrm{T}})$ denotes the position of the Tx whose PDF is $g_{\mathbf{r}_{\mathrm{T}}}(\mathbf{r})$, 
{\color{black}$f_{\tau_\mathrm{I},\chi|d,\kappa,\upsilon,\mathbf{r}_{\mathrm{T}}}(\tau_{\mathrm{I}})$} is the PDF of RMS-DS with given scenario $\upsilon$ and blockage condition $\kappa$, and $\xi_i$ denotes the event that Tx is located in the $i$-th room. Since $f_{\tau_\mathrm{I},\chi|d,\kappa,\upsilon,\mathbf{r}_{\mathrm{T}}}(\tau_{\mathrm{I}})$ denotes the PDF of $\tau$, it requires specific distribution assumptions of $\tau$ to be further determined.}
Given that the BUD is composed of $N_{\mathrm{r}}$ rooms, and the $i$-th room is with type $\upsilon_i$ and an area of $S_i$. Then the total area of the entire building is calculated by 
\begin{equation}
\label{V_general}
V=\sum_{i=1}^{N_{\mathrm{r}}}S_i.
\end{equation} 
Then $\mathrm{E}[\tau_{\mathrm{I}}]$ can be calculated as (\ref{eq5}).


\begin{figure*}[b]
	
{\color{black}	\begin{eqnarray}
	\label{eq5}
	\mathrm{E}[\tau_{\mathrm{I}}]
	=\int_0^{+\infty}\tau_{\mathrm{I}}
	\sum_{i=1}^{N_\mathrm{r}}
	\int_0^{+\infty}p_{\iota,\chi}(d)
	\int_{\mathbf{r}}
	\sum_{\kappa}\sum_{\upsilon}f_{\tau_\mathrm{I},\chi|d,\kappa,\upsilon_i,\mathbf{r}_{\mathrm{T}}}(\tau)\mathcal{P}_{\mathbf{r}_\mathrm{T}}(\upsilon)\mathcal{P}_{\xi_i,d}(\kappa|\chi)
	g_{\mathbf{r}_{\mathrm{T}}|\xi_i}(\mathbf{r})
	\mathcal{P}(\xi_i) \mathrm{d}\mathbf{r}\mathrm{d}d\mathrm{d}\tau_\mathrm{I}.
	\end{eqnarray}}
\end{figure*}

The aim of this section is to provide an analytical evaluation of the DS gain for a BUD. To facilitate the derivation, we make the following assumptions.
\begin{itemize}
	\item All rooms in the BUD are assumed to be rectangular. The size of the $i$-th room $S_i$ can be calculated by $m_i\times l_i$, where $m_i$ denotes the long edge and $l_i$ denotes the short edge of the $i$-th room. The BUD is assumed to be rectangular as well, with a size of $X\times Y$, where $X$ denotes the long edge of the BUD, and $Y$ denotes its short edge.  We set the coordinate system on BUD, with origin at the left bottom.
	With this assumption, \eqref{V_general} can be extended to:
	\begin{equation}
	\label{V_assume}
	V=\sum_{i=1}^{N_{\mathrm{r}}}S_i=XY.
	\end{equation} 
	\item Both the Tx and the Rx are assumed to be distributed in the BUD uniformly with heights $h_\mathrm{T}$ and $h_\mathrm{R}$. Therefore, 
	\begin{eqnarray}
	\label{eq10}
	g_{\mathbf{r}_{\mathrm{T}}}(\mathbf{r})=\left\{
	\begin{array}{ll}
	\frac{1}{V}& 
	\mathbf{r}_\mathrm{T} \in  \mathcal{C}_\mathrm{B},
	\\
	0,& \mathrm{else},
	\end{array}
	\right.
	\end{eqnarray}
	and 
	\begin{eqnarray}
	\label{eq8}
	\mathcal{P}(\xi_i)=\frac{S_i}{V},
	\end{eqnarray}
	where $\mathcal{C}_\mathrm{B}$ is defined as the set of coordinates occupied by the BUD.
	Similarly, we define the set of coordinates occupied by the $i$-th room as $\mathcal{C}_i$. Therefore, the conditional PDF of $\mathbf{r}_{\mathrm{T}}$ when Tx is located in the $i$-th room is defined as $g_{\mathbf{r}_{\mathrm{T}}|\xi_i}(\mathbf{r})$ :
	\begin{eqnarray}
		\label{eq9p5}
		g_{\mathbf{r}_{\mathrm{T}}|\xi_i}(\mathbf{r})=\left\{
		\begin{array}{ll}
		\frac{1}{S_i}& 
		\mathbf{r}_\mathrm{T}\in\mathcal{C}_i,
		\\
		0,& \mathrm{else}.
		\end{array}
		\right.
	\end{eqnarray}
	\item We assume that two possible types of rooms, which are offices and corridors, consist of the BUD,  i.e., $\upsilon=\{\mathrm{office},\mathrm{corridor}\}$. We define the sum size of offices as $S_\mathrm{offi}$ and corridors as $S_\mathrm{corr}$, and define the set of coordinates occupied by the offices as $\mathcal{C}_\mathrm{offi}$  and the corridors as $\mathcal{C}_\mathrm{corr}$.
	Therefore,
	\begin{eqnarray}
	\label{eq10}
	\mathcal{P}_{\mathbf{r}_\mathrm{T}}(\upsilon)=\left\{
	\begin{array}{ll}
	\frac{S_\mathrm{offi}}{V}& 
	\mathbf{r}_\mathrm{T} \in  \mathcal{C}_\mathrm{offi},\\
	\frac{S_\mathrm{corr}}{V}& 
	\mathbf{r}_\mathrm{T} \in  \mathcal{C}_\mathrm{corr},\\
	0 &\mathrm{else}
	\end{array}
	\right.
	\end{eqnarray}
	
	\begin{table}[!t]  
		\centering  
		\scriptsize  
		\caption{Parameters of the RMS-DS model}  
		\label{tab:parameters}  
		\begin{tabular}{ c c c c c c c c}  
			\toprule
			$\upsilon$& $\kappa$ & $k$& $B$& $\sigma$& $n$& $C$& $\sigma_s$\\  
			\midrule  
			\multirow{2}{*}{Office}      &   LOS& 0.40 & -3.43 & 2.34 & 2.55 & 0.37 & 3.76\\
			& NLOS & 0.40 & -4.77 & 3.30 & 2.40 & 10.73 & 3.62\\
			\multirow{2}{*}{Corridor}      &   LOS& 0.38 & -5.72 & 2.40 & 1.81 & 0.32 & 2.69\\
			& NLOS & 0.39 & -8.04 & 2.97 & 1.82 & 5.56 & 2.73\\
			
			\bottomrule 
		\end{tabular}  
	\end{table}

	\item We assume that RMS-DS has a linear relationship with PL in indoor environments according to measurement results in \cite{yu2017measurement}. More specifically, RMS-DS can be expressed as follows:
{\color{black}	\begin{equation}
	\label{eq11}
	\tau_{\kappa,\upsilon,d,\mathrm{I}}=k_{\kappa,\upsilon}{L}_{\kappa,\upsilon,\mathrm{I}}(d)+B_{\kappa,\upsilon}+X_{\kappa,\upsilon,\mathrm{z}},
	\end{equation}}
	where $X_{\kappa,\upsilon,\mathrm{z}}$ is a normally distributed random variable with the mean value $\mu_{\kappa,\upsilon}=0$ and standard deviation $\sigma_{\kappa,\upsilon}$. {\color{black}${L}_{\kappa,\upsilon,\mathrm{I}}(d)$ }is the path loss which can be described as:
	{\color{black}\begin{equation}
	\label{eq12}
	{L}_{\kappa,\upsilon,\mathrm{I}}(d)={L}(d_0)+10n_{\kappa,\upsilon}\mathrm{log}_{10}(d/d_0)+C_{\kappa,\upsilon}+X_{\kappa,\upsilon,s},
	\end{equation}}where $L(d_0)$ is the reference PL in decibel ($d_0=1$ m and the test band is centered at $2.595$ GHz), which is $40.7$ dB according to \cite{yu2017measurement}. $X_{\kappa,\upsilon,s}$ is a normally distributed random variable with the mean value $\mu_{\kappa,\upsilon,s}=0$ and standard deviation $\sigma_{\kappa,\upsilon,s}$.
	Values of parameters $k_{\kappa,\upsilon}$, $B_{\kappa,\upsilon}$, $\sigma_{\kappa,\upsilon}$, $n_{\kappa,\upsilon}$, $C_{\kappa,\upsilon}$ and $\sigma_{\kappa,\upsilon,s}$ vary according to the $\kappa$ and $\upsilon$. In this paper, to validate the RMS-DS expectation model, the parameter values for different $\kappa$ and $\upsilon$ in \cite{yu2017measurement} are employed, as shown in Table \ref{tab:parameters}. The model suggested in this paper is still valid for the room types not mentioned, as long as the necessary experimental data is supplied. 
	In addition, in this paper, the RMS-DS in open space is calculated using a two-ray model. 
	
	By combining Eqs. (\ref{eq11}) and (\ref{eq12}), for Tx locating at the $i$-th room, the RMS-DS of the indoor channel can be represented as a normal distributed random variable with the expectation value: 
{\color{black}	\begin{equation}
	 \begin{split}
	\label{RMS:mu}
	\mu_{\kappa,\upsilon_i,\tau_\mathrm{I}}(d)&=k_{\kappa,\upsilon_i}L(d_0)+10k_{\kappa,\upsilon_i}n_{\kappa,\upsilon_i}\mathrm{log}_{10}(d/d_0)\\
	&+k_{\kappa,\upsilon_i}C_{\kappa,\upsilon_i}+B_{\kappa,\upsilon_i},
	 \end{split}
	\end{equation}}
	and the standard deviation:
	{\color{black}\begin{equation}
	\label{RMS:sigma}
	\sigma_{\kappa,\upsilon_i,\tau_\mathrm{I}}=\sqrt{\sigma^2_{\kappa,\upsilon_i}+{k_{\kappa,\upsilon_i}^2\sigma^2_{\kappa,\upsilon_i,s}}}.
	\end{equation}}
	
	Therefore, {\color{black}$f_{\tau_\mathrm{I},\chi|d,\kappa,\upsilon_i,\mathbf{r}_{\mathrm{T}}}(\tau_\mathrm{I})$ }is given by 
	

{\color{black}	\begin{equation}
	\label{eq15}
	f_{\tau_\mathrm{I},\chi|d,\kappa,\upsilon_i,\mathbf{r}_{\mathrm{T}}}(\tau_\mathrm{I})=\frac{\mathrm{exp}\left(-\frac{1}{2}\left(\frac{\tau_\mathrm{I}-\mu_{\kappa_i,\upsilon,\tau_\mathrm{I}}(d)}{\sigma_{\kappa_i,\upsilon,\tau_\mathrm{I}}}\right)^2\right)}{\sqrt{2\pi}\sigma_{\kappa_i,\upsilon,\tau_\mathrm{I}}} 
	\end{equation}}
	under this assumption.
\end{itemize}

From (\ref{eq5}), we can see that with given (\ref{eq8}), (\ref{eq9p5}), \eqref{eq10}, and (\ref{eq15}), as long as we have the analytical expression of $\mathcal{P}_{\xi_i,d}(\kappa|\chi)$ and $
p_{\iota,\chi}(d)$, $\mathrm{E}[\tau_{\mathrm{I}}]$ can be derived analytically.
$\mathcal{P}_{\xi_i,d}(\kappa|\chi)$ and $
p_{\iota,\chi}(d)$ are given by the following Lemma 1 and Lemma 2, respectively.

\begin{lemma}

	\begin{eqnarray}
\label{P_v|d}
\mathcal{P}_{\xi_i,d}(\kappa|\chi)=\left\{
\begin{array}{ll}
\frac{Z(d,l_i,m_i)}{Z(d,Y,X)},& 
\mathrm{LOS},
\\
1-\frac{Z(d,l_i,m_i)}{Z(d,Y,X)},& \mathrm{NLOS},
\end{array}
\right.
\end{eqnarray}
	where $Z(d,a,b), (a<b)$ is defined as \eqref{Z_d}.
\begin{figure*}
	\begin{equation}
	\label{Z_d}
	Z(d,a,b)=\left\{
	\begin{array}{lcl}
	\frac{d^2-2d(a+b)+ab\pi}{ab\pi}, & & {0< d\leq a},\\
	\frac{-a^2+2db(\sqrt{1-\frac{l^2}{d^2}}-1)+2ab\mathrm{arcsin}\frac{a}{d}}{ab\pi}, & & {a< d\leq b},\\
	-\frac{d^2+a^2+b^2-2d(b\sqrt{1-\frac{a^2}{d^2}}+a\sqrt{1-\frac{b^2}{d^2}})+2ab(\mathrm{arccos}\frac{a}{d}+\mathrm{arccos}\frac{b}{d}-\frac{\pi}{2})}{ab\pi}, & & {b< d\leq \sqrt{a^2+b^2}},\\
	0, & & {\sqrt{a^2+b^2}}< d.
	\end{array}\right.
	\end{equation}
\end{figure*}
\end{lemma}
\begin{IEEEproof}
	{\color{black}The function $Z(d,a,b)$ is defined for:
		In the promise that a point A is in a rectangle with the size $a\times b, (a<b)$, the probability that a point B at a distance $d$ from the point A is also in the rectangle is calculated by $Z(d,a,b)$ \cite[Eq. (3)]{zheng2018exact}.	

	Mimicking the function $Z(d,a,b)$, we define that when Tx is in the $i$-th room and the Tx-Rx distance is $d$, the probability that the Tx-Rx link is LOS is $\mathcal{P}_{\xi_i,d}(\mathrm{LOS})$. Similarly, $\mathcal{P}_{\xi_i,d}(\chi)$ denotes the probability that Rx is located in the BUD when Tx is in the $i$-th room, and the distance between Tx and RX is $d$.}
	According to the definition of $Z(d,a,b)$, we can derive:
	\begin{eqnarray}
		\label{P_los_room}
		\mathcal{P}_{\xi_i,d}(\mathrm{LOS})=Z(d,l_i,m_i),
	\end{eqnarray}
and 
	\begin{equation}
	\label{P_los_BUD}
	\mathcal{P}_{\xi_i,d}(\chi)=Z(d,Y,X),
	\end{equation}
	
	{\color{black}When the Tx-Rx link is LOS and Tx is located in BUD, then Rx is always located in BUD.
	Therefore, we can derive:
	\begin{equation}
	\label{P_los_BUD}
	\begin{split}
	\mathcal{P}_{\xi_i,d}(\mathrm{LOS}|\chi)=\frac{\mathcal{P}_{\xi_i,d}(\mathrm{LOS},\chi)}{\mathcal{P}_{\xi_i|d}(\chi)}\\
	=\frac{\mathcal{P}_{\xi_i,d}(\mathrm{LOS})}{\mathcal{P}_{\xi_i|d}(\chi)}
	=\frac{Z(d,l_i,m_i)}{Z(d,Y,X)},
	\end{split}
	\end{equation}
	and therefore}
	\begin{equation}
	\label{P_nlos_BUD}
	\begin{split}
	\mathcal{P}_{\xi_i,d}&(\mathrm{NLOS}|\chi)=1-\mathcal{P}_{\xi_i,\chi}(\mathrm{LOS}|\chi)\\
	&=1-\frac{Z(d,l_i,m_i)}{Z(d,Y,X)}.
	\end{split}
	\end{equation}
\end{IEEEproof}

\begin{lemma}
	
	\begin{eqnarray}
	\label{p(d)}
	p_{\iota,\chi}(d)= \frac{2\pi d Z(d,Y,X)}{XY},
	\end{eqnarray}
	where $Z(d)$ is defined as (\ref{Z_d}) as well.
\end{lemma}
 \begin{figure}[!t]
	\centering
	\includegraphics [width=3in]{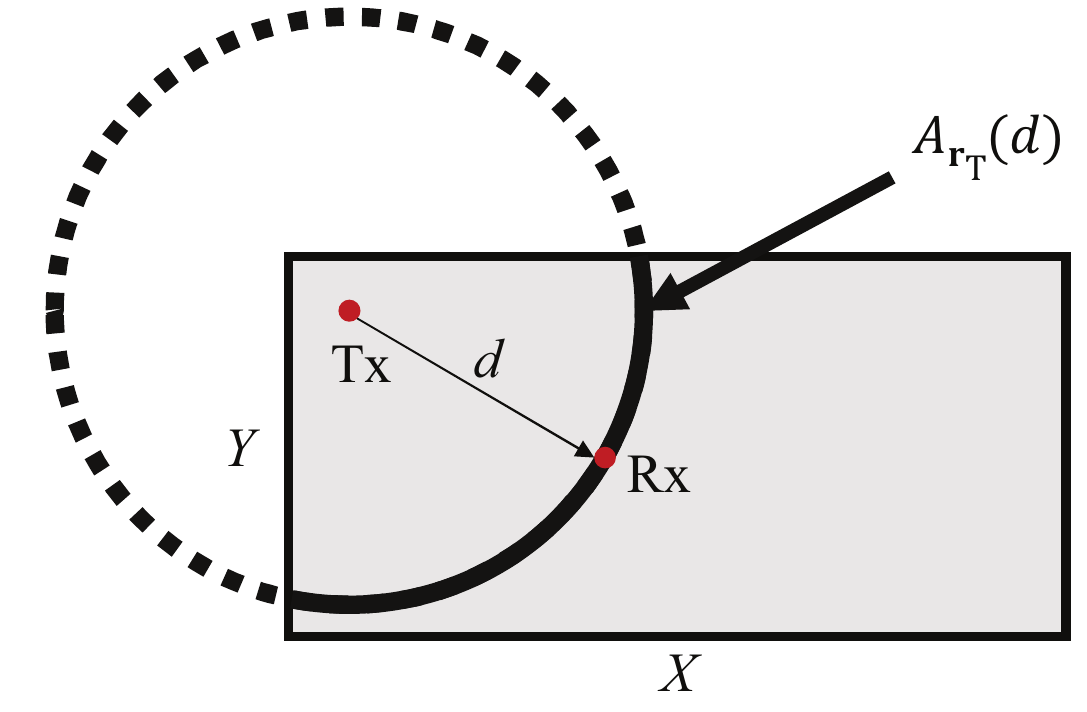}
	\caption{For a transmitter Tx at any position in the building, when the distance between it and the receiver Rx is $d$, Rx is located on an arc of radius $d$ with Tx as the center of the circle.}
	\label{distance_prob}
	\centering
\end{figure}
\begin{IEEEproof}
	{\color{black}As Fig. \ref{distance_prob} shows, for Tx locating at $\mathbf{r}_{\mathrm{T}}$ in the building,  all possible positions of receiver Rx which is $d$ far from Tx, can form a circle with the center of $\mathbf{r}_{\mathrm{T}}=(x_\mathrm{T},y_\mathrm{T})$ and radius of  $d$, whose circumference is $2\pi d$.
	Rx that locates on the arc of the circle in the building can communicate with Tx in LOS condition. We define the length of the arc inside the building as $A_{\mathbf{r}_{\mathrm{T}}}(d)$.
	Under this premise, when Rx is randomly located at any position in an infinitely extended plane, the probability that it locates on $A_{\mathbf{r}_{\mathrm{T}}}(d)$ is defined as $\mathcal{P}_{\mathrm{arc},\mathbf{r}_{\mathrm{T}}}(d)$. Similarly, we define the probability that an Rx locates in the BUD as $\mathcal{P}_{\chi,\mathbf{r}_{\mathrm{T}}}(d)$.
	
	Therefore, for any Rx in the building, the probability that its distance from Tx locating at $\mathbf{r}_{\mathrm{T}}$ is $d$ can be defined as $\mathcal{P}_{\mathbf{r}_{\mathrm{T}}}(d|\chi)$, which can be calculated as:
	\begin{equation}
	\label{arc}
	\mathcal{P}_{\mathbf{r}_{\mathrm{T}}}(d|\chi)= \frac{\mathcal{P}_{\mathrm{arc},\mathbf{r}_{\mathrm{T}}}(d)}{\mathcal{P}_{\chi,\mathbf{r}_{\mathrm{T}}}(d)}=\frac{A_{\mathbf{r}_{\mathrm{T}}}(d)/\Phi}{XY/\Phi}=\frac{A_{\mathbf{r}_{\mathrm{T}}}(d)}{XY},
	\end{equation}
	where $\Phi$ denotes the size of the infinitely extended plane.
	
	We define the probability that distance between an Rx with Tx locating at $\mathbf{r}_{\mathrm{T}}$ is $d$ as $\mathcal{P}_{\mathbf{r}_{\mathrm{T}}}(d)$.
	With given Tx locating at $\mathbf{r}_{\mathrm{T}}$, the probability that an Rx is inside the building when it is $d$ far from Tx can be defined as $\mathcal{P}_{\mathbf{r}_\mathrm{T}}(\chi|d)$, which can be calculated as:}
	\begin{equation}
	\label{P_LOS_xy}
	\mathcal{P}_{\mathbf{r}_\mathrm{T}}(\chi|d)=\frac{\mathcal{P}_{\mathrm{arc},\mathbf{r}_\mathrm{T}}(d)}{\mathcal{P}_{\mathbf{r}_{\mathrm{T}}}(d)}= \frac{A_{\mathbf{r}_{\mathrm{T}}}(d)}{2\pi d}.
	\end{equation}
%
	
	Combining Eq. \eqref{arc} and \eqref{P_LOS_xy}, we can derive:
	\begin{equation}
	\label{eq26}
	\begin{split}
	\mathcal{P}_{\mathbf{r}_{\mathrm{T}}}(d|\chi)= \frac{2\pi d \mathcal{P}_{\mathbf{r}_\mathrm{T}}(\chi|d)}{XY}.
	\end{split}
	\end{equation}
	
	{\color{black}For Tx at all possible positions in the BUD, the probability that the randomly located Rx is $d$ far from Tx is in the building is:}
	\begin{equation}
	\label{eq27}
	\begin{split}
	\mathcal{P}(\chi|d)=\int_{0}^{Y}\int_{0}^{X} \mathcal{P}_{\mathbf{r}_\mathrm{T}}(\chi|d)\mathrm{d}x_\mathrm{T}\mathrm{d}y_\mathrm{T}.
	\end{split}
	\end{equation}
	{\color{black}Based on the definition of $Z(d,a,b)$ that has been explained in Lemma 1, \eqref{P_LOS_sum} can be expressed as:
    \begin{equation}
	\label{eq28}
	\begin{split}
	\mathcal{P}(\chi|d)=Z(d,Y,X).
	\end{split}
	\end{equation}
	
	Combining \eqref{eq26}, \eqref{eq27} and \eqref{eq28}, for any randomly located Rx in the building, the probability that its distance from Tx located at $\mathbf{r}_\mathrm{T}$ is $d$ can be derived as:}
	\begin{equation}
	\label{P_LOS_sum}
	\begin{split}
	p_{\iota,\chi}(d)= \int_{0}^{Y}\int_{0}^{X} \mathcal{P}_{\mathbf{r}_\mathrm{T}}(d|\chi)\mathrm{d}x_\mathrm{T}\mathrm{d}y_\mathrm{T}\\
	=\int_{0}^{Y}\int_{0}^{X} \frac{2\pi d \mathcal{P}_{\mathbf{r}_\mathrm{T}}(\chi|d)}{XY}\mathrm{d}x_\mathrm{T}\mathrm{d}y_\mathrm{T}\\
	=\frac{2\pi d \mathcal{P}(\chi|d)}{XY}=\frac{2\pi d Z(d,Y,X)}{XY}.
	\end{split}
	\end{equation}
\end{IEEEproof}

\begin{theorem}
	The $\mathrm{E}[\tau_{\mathrm{I}}]$ can be derived as an analytical expression as (\ref{f_I(tau)_simp}).
\begin{figure*}
	{\color{black}
	\begin{align}
	&\mathrm{E}[\tau_{\mathrm{I}}]
	=\int_0^{+\infty}\tau_\mathrm{I}\int_0^{+\infty}
	\sum_{i}^{N_\mathrm{r}}\frac{S_i}{V}\frac{2\pi d }{XY}\left[\frac{Z(d,l_i,m_i)}{\sqrt{2\pi}\sigma_{\mathrm{LOS},\upsilon_i,\tau_\mathrm{I}}} \mathrm{exp}\left(-\frac{1}{2}\left(\frac{\tau_\mathrm{I}-\mu_{\mathrm{LOS},\upsilon_i,\tau_\mathrm{I}}(d)}{\sigma_{\mathrm{LOS},\upsilon_i,\tau_\mathrm{I}}}\right)^2\right)
	\right.\nonumber\\
	\label{f_I(tau)}
	&\left.+\frac{Z(d,Y,X)-Z(d,l_i,m_i)}{\sqrt{2\pi}\sigma_{\mathrm{NLOS},\upsilon_i,\tau_\mathrm{I}}} \mathrm{exp}\left(-\frac{1}{2}\left(\frac{\tau_\mathrm{I}-\mu_{\mathrm{NLOS},\upsilon_i,\tau_\mathrm{I}}(d)}{\sigma_{\mathrm{NLOS},\upsilon_i,\tau_\mathrm{I}}}\right)^2\right)\right]\mathrm{d}d\mathrm{d}\tau_\mathrm{I},\\
	&=\sum_{i}^{N_\mathrm{r}}\frac{S_i}{V}\int_{0}^{+\infty}\frac{2\pi d}{XY}\left[Z(d,l_i,m_i)\left(\frac{\sigma_{\mathrm{LOS},\upsilon_i,\tau_\mathrm{I}}}{\sqrt{2\pi}}\mathrm{exp}\left(-\frac{1}{2}\left(\frac{\mu_{\mathrm{LOS},\upsilon_i,\tau_\mathrm{I}}(d)}{\sigma_{\mathrm{LOS},\upsilon_i,\tau_\mathrm{I}}}\right)^2\right)+\frac{\mu_{\mathrm{LOS},\upsilon_i,\tau_\mathrm{I}}(d)}{2}\mathrm{erfc}\left(-\frac{\mu_{\mathrm{LOS},\upsilon_i,\tau_\mathrm{I}}(d)}{\sqrt{2}\sigma_{\mathrm{LOS},\upsilon_i,\tau_\mathrm{I}}}\right)\right)\right.\nonumber \\
	&\left.+\left(Z\left(d,Y,X\right)-Z\left(d,l_i,m_i\right)\right)\left(\frac{\sigma_{\mathrm{NLOS},\upsilon_i,\tau_\mathrm{I}}}{\sqrt{2\pi}}\mathrm{exp}\left(-\frac{1}{2}\left(\frac{\mu_{\mathrm{NLOS},\upsilon_i,\tau_\mathrm{I}}(d)}{\sigma_{\mathrm{NLOS},\upsilon_i,\tau_\mathrm{I}}}\right)^2\right)+\frac{\mu_{\mathrm{NLOS},\upsilon_i,\tau_\mathrm{I}}(d)}{2}\mathrm{erfc}\left(-\frac{\mu_{\mathrm{NLOS},\upsilon_i,\tau_\mathrm{I}}(d)}{\sqrt{2}\sigma_{\mathrm{NLOS},\upsilon_i,\tau_\mathrm{I}}}\right)\right) \right]\mathrm{d}d. \label{f_I(tau)_simp}
	\end{align}}
\end{figure*}
\end{theorem}
\begin{IEEEproof}
	By substituting \eqref{eq8}, \eqref{eq9p5}, \eqref{eq10}, \eqref{eq15}, \eqref{P_v|d}, and \eqref{p(d)} into (\ref{eq5}), we have an analytical expression of $\mathrm{E}[\tau_{\mathrm{I}}]$ as (\ref{f_I(tau)}) via straightforward derivations.
	After exchanging the integral order and some straight forward derivation, we can obtain a more computationally tractable analytical solution as (\ref{f_I(tau)_simp}).
\end{IEEEproof}

\subsection{Derivation of $\mathrm{E}[\tau_{\mathrm{O}}]$}
With the same assumption in subsection III-A, $\mathrm{E}[\tau_{\mathrm{O}}]$ can be calculated as (\ref{tau_O_orin}) within the open space whose range is the same as the BUD.
\begin{equation}
\label{tau_O_orin}
\begin{split}
\mathrm{E}[\tau_{\mathrm{O}}]= \int_0^{+\infty}p_{\iota,\chi}(d)\int_\mathbf{r} f_{\tau_\mathrm{O},\chi|\mathbf{r}_{\mathrm{T}}}(d) g_{\mathbf{r}_{\mathrm{T}}}(\mathbf{r}) \mathrm{d}\mathbf{r}\mathrm{d}d.
\end{split}
\end{equation}
where $f_{\tau_\mathrm{O},\chi|\mathbf{r}_{\mathrm{T}}}(d)$ denotes the RMS-DS in the open space when the transmission distance is $d$.
Given \eqref{eq8}, \eqref{eq9p5}, and (\ref{p(d)}), as long as we have the analytical expression of $f_{\tau_\mathrm{O},\chi|\mathbf{r}_{\mathrm{T}}}(d)$, we can derive $\mathrm{E}[\tau_{\mathrm{O}}]$. $f_{\tau_\mathrm{O},\chi|\mathbf{r}_{\mathrm{T}}}(d)$ is given in Lemma 3.
\begin{lemma}
	$f_{\tau_\mathrm{O},\chi|\mathbf{r}_{\mathrm{T}}}(d)$ can be derived in a closed-form expression as \eqref{f_O(d)}, where $c$ denotes the light speed.
\end{lemma}

\begin{IEEEproof}
In this paper, we have assumed that the RMS-DS in the open space is calculated using a two-ray model, {\color{black}as fig. \ref{2raymodel} shows. In the two-ray model, the RMS-DS is given by the root mean square of the delay of the directive path and the delay of the reflective path as \eqref{2ray}.} Therefore, $f_{\tau_\mathrm{O},\chi|\mathbf{r}_{\mathrm{T}}}(d)$ can be derived as \eqref{f_O(d)}, where $c$ denotes the light speed.

 \begin{figure}[!t]
	\centering
	\includegraphics [width=3in]{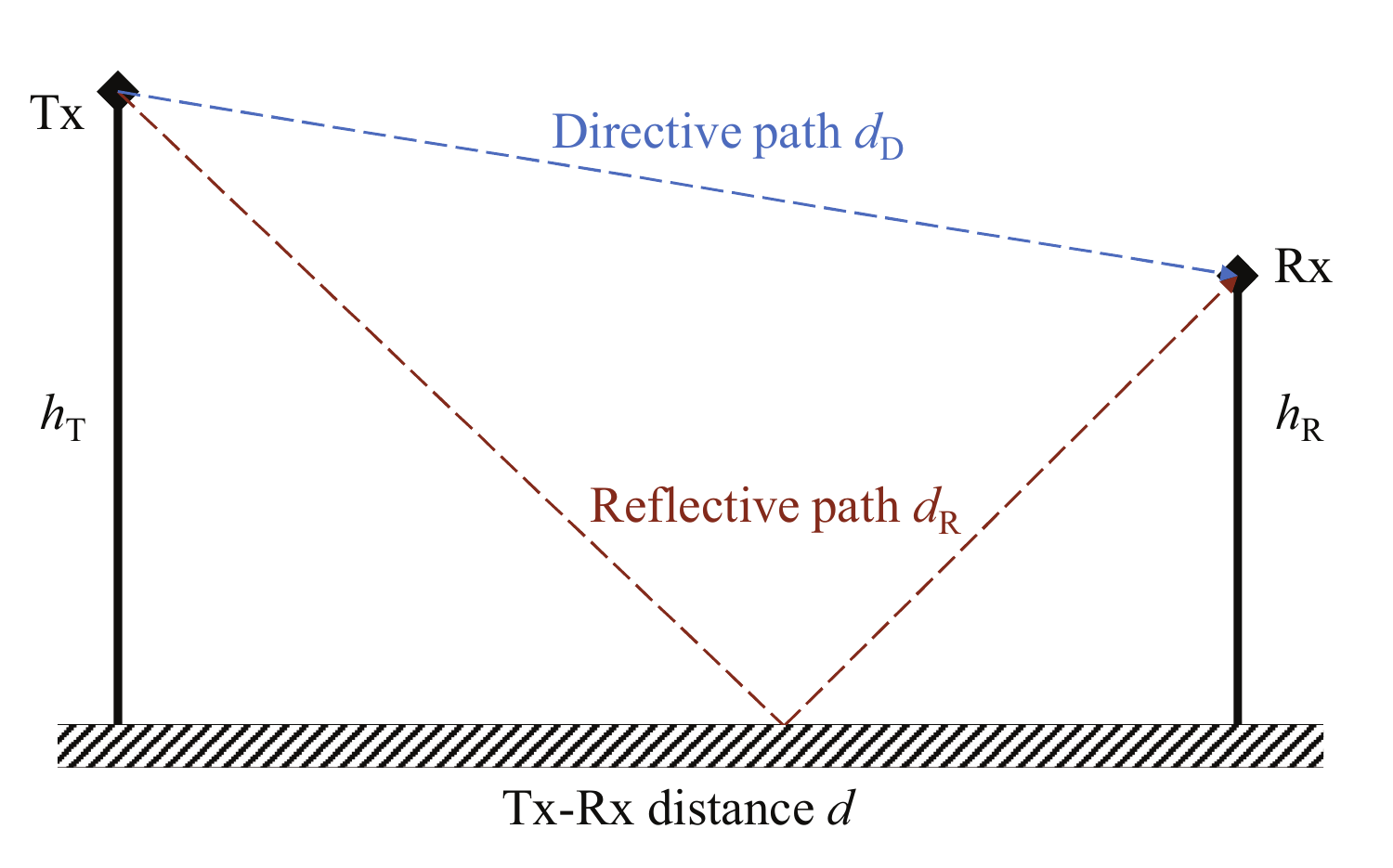}
	\caption{Two ray model.}
	\label{2raymodel}
	\centering
\end{figure}

\begin{figure*}[t!]
{\color{black}	\begin{align}
	f_{\tau_\mathrm{O},\chi|\mathbf{r}_{\mathrm{T}}}(d)&=\sqrt{\frac{\left(\frac{d_\mathrm{D}}{c}-\frac{d_\mathrm{D}+d_\mathrm{R}}{2c}\right)^2+\left(\frac{d_\mathrm{R}}{c}-\frac{d_\mathrm{D}+d_\mathrm{R}}{2c}\right)^2}{2}} \label{2ray}
	\\
	&=\frac{1}{2c}\left|\sqrt{d^2+{(h_\mathrm{t}-h_\mathrm{r})}^2}-\sqrt{h_\mathrm{t}^2+\left(\frac{dh_\mathrm{t}}{h_\mathrm{t}+h_\mathrm{r}}\right)^2}-\sqrt{h_\mathrm{r}^2+\left(\frac{dh_\mathrm{r}}{h_\mathrm{t}+h_\mathrm{r}}\right)^2} \right|. \label{f_O(d)}
	\end{align}}
\end{figure*}

\end{IEEEproof}

\begin{theorem}
	The $\mathrm{E}[\tau_{\mathrm{O}}]$ can be derived in an analytical expression as (\ref{tau_O}).
\begin{figure*}[t!]
	
	\begin{equation}
	\label{tau_O}
	\begin{split}
	\mathrm{E}[\tau_{\mathrm{O}}]=\int_0^{+\infty}\frac{\pi d Z(d,Y,X)}{XYc}
	\left|\sqrt{d^2+{(h_\mathrm{t}-h_\mathrm{r})}^2}-\sqrt{h_\mathrm{t}^2+\left(\frac{dh_\mathrm{t}}{h_\mathrm{t}+h_\mathrm{r}}\right)^2}-\sqrt{h_\mathrm{r}^2+\left(\frac{dh_\mathrm{r}}{h_\mathrm{t}+h_\mathrm{r}}\right)^2} \right|\mathrm{d}d.
	\end{split}
	\end{equation}
\end{figure*}
	
\end{theorem}
\begin{IEEEproof}
	By sustituting \eqref{eq8}, \eqref{eq9p5}, (\ref{p(d)}), and (\ref{f_O(d)}) into (\ref{tau_O_orin}), the analytical expression of $\mathrm{E}[\tau_{\mathrm{O}}]$ can be obtained as (\ref{tau_O}) via straightforward derivations.
\end{IEEEproof}

\subsection{Derivation of $G_\tau$}

Based on Theorems 1 and 2, we can derive the analytic expression of $G_\tau$ as follows. 

By substituting (\ref{f_I(tau)_simp}) and (\ref{tau_O}) into (\ref{gderive}), 
we finally obtain the analytical expression of $G_\tau$ in \eqref{eq34}.

\begin{figure*}[t!]
{\color{black}\begin{align}
&G_\tau =\sum_{i}^{N_\mathrm{r}}\frac{S_i}{V}\int_{0}^{+\infty}\frac{2\pi d}{XY}\left[Z(d,l_i,m_i)\left(\frac{\sigma_{\mathrm{LOS},\upsilon_i,\tau_\mathrm{I}}}{\sqrt{2\pi}}\mathrm{exp}\left(-\frac{1}{2}\left(\frac{\mu_{\mathrm{LOS},\upsilon_i,\tau_\mathrm{I}}(d)}{\sigma_{\mathrm{LOS},\upsilon_i,\tau_\mathrm{I}}}\right)^2\right)+\frac{\mu_{\mathrm{LOS},\upsilon_i,\tau_\mathrm{I}}(d)}{2}\mathrm{erfc}\left(-\frac{\mu_{\mathrm{LOS},\upsilon_i,\tau_\mathrm{I}}(d)}{\sqrt{2}\sigma_{\mathrm{LOS},\upsilon_i,\tau_\mathrm{I}}}\right)\right)\right.\nonumber\\
&\left.+\left(Z(d,Y,X)-Z(d,l_i,m_i)\right)\left(\frac{\sigma_{\mathrm{NLOS},\upsilon_i,\tau_\mathrm{I}}}{\sqrt{2\pi}}\mathrm{exp}\left(-\frac{1}{2}\left(\frac{\mu_{\mathrm{NLOS},\upsilon_i,\tau_\mathrm{I}}(d)}{\sigma_{\mathrm{NLOS},\upsilon_i,\tau_\mathrm{I}}}\right)^2\right)+\frac{\mu_{\mathrm{NLOS},\upsilon_i,\tau_\mathrm{I}}(d)}{2}\mathrm{erfc}\left(-\frac{\mu_{\mathrm{NLOS},\upsilon_i,\tau_\mathrm{I}}(d)}{\sqrt{2}\sigma_{\mathrm{NLOS},\upsilon_i,\tau_\mathrm{I}}}\right)\right) \right]   \mathrm{d}d \nonumber\\
&-\int_0^{+\infty} \frac{\pi d Z(d,Y,X)}{XYc}
\left|\sqrt{d^2+{(h_\mathrm{t}-h_\mathrm{r})}^2}-\sqrt{h_\mathrm{t}^2+\left(\frac{dh_\mathrm{t}}{h_\mathrm{t}+h_\mathrm{r}}\right)^2}-\sqrt{h_\mathrm{r}^2+\left(\frac{dh_\mathrm{r}}{h_\mathrm{t}+h_\mathrm{r}}\right)^2} \right| \mathrm{d}d \label{eq34}.
\end{align}}
\end{figure*}

{\color{black}
\subsection{Model Reliability}
We propose a metric $\sigma$ to assess the model's reliability to assist architects in better using this evaluation model. We analyze the standard deviation between the analytical and estimated DS gain of any Tx-Rx link in BUD and then derive $\sigma$ by calculating the expectation value of the standard deviation of all links in BUD in accordance with the distance distribution $p_{\iota,\chi}(d)$. The mathematical expression of $\sigma$ is as \eqref{sigma_math} shows.

Since $f_{\tau_\mathrm{O}}(d)$ does not randomly vary, \eqref{sigma_math} can be simplified to \eqref{sigma_2}.
Additionally, as $f_{\tau_\mathrm{O}}(d)$ is weighted by different Gaussian distributions according to \eqref{eq5p5}, \eqref{sigma_2} can be further deduced as \eqref{sigma_3} using the nature of Gaussian distributions. Therefore, $\sigma$ can be finally deduced as an analytical expression as \eqref{sigma_4}.
}

\begin{figure*}[t!]
{\color{black} \begin{align}
\sigma &=\int_0^{+\infty} p_{\iota,\chi}(d) \sqrt{ \mathbb{E}\left[\left(\left(f_{\tau_\mathrm{I}}(d)-f_{\tau_\mathrm{O}}(d)\right)-\left(\int_0^{+\infty}\tau f_{\tau_\mathrm{I}}(d)\mathrm{d}\tau-f_{\tau_\mathrm{O}}(d)\right)\right)^2\right]}\mathrm{d}d \label{sigma_math}\\
&=\int_0^{+\infty} p_{\iota,\chi}(d) \sqrt{ \mathbb{E}\left[\left(f_{\tau_\mathrm{I}}(d)-\int_0^{+\infty}\tau f_{\tau_\mathrm{I}}(d)\mathrm{d}\tau\right)^2\right]}\mathrm{d}d 
=\int_0^{+\infty} p_{\iota,\chi}(d) \sqrt{ \mathbb{D}\left[f_{\tau_\mathrm{I}}(d)\right]}\mathrm{d}d \label{sigma_2}\\
&=\int_0^{+\infty} p_{\iota,\chi}(d) \sqrt{ \sum_i^{N_\mathrm{r}} \sum_{\kappa} \mathcal{P}_{\xi_i,d}(\kappa|\chi) \sigma_{\kappa,\upsilon_i,\tau_\mathrm{I}}}	\mathrm{d}d \label{sigma_3}\\
&=\int_0^{+\infty} \frac{2\pi d Z(d,Y,X)}{XY} \sqrt{ \sum_i^{N_\mathrm{r}} \left[ \frac{Z(d,l_i,m_i) }{Z(d,l_i,m_i)}\sigma_{\mathrm{LOS},\upsilon_i,\tau_\mathrm{I}}+\left(1-\frac{Z(d,l_i,m_i) }{Z(d,l_i,m_i)}\right)\sigma_{\mathrm{NLOS},\upsilon_i,\tau_\mathrm{I}}
\right]}	\mathrm{d}d \label{sigma_4}
\end{align}}
\end{figure*}

\begin{figure}[!t]
	\centering
	\subfigure[2 by 3]{
		\includegraphics[width=1.27in]{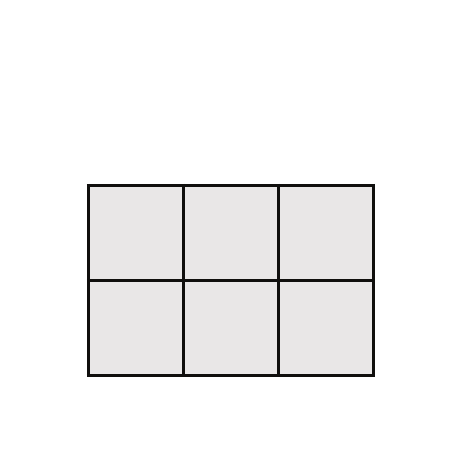}\label{2B3}}\hspace{-8.5mm}
	\subfigure[3 by 3]{
		\includegraphics[width=1.27in]{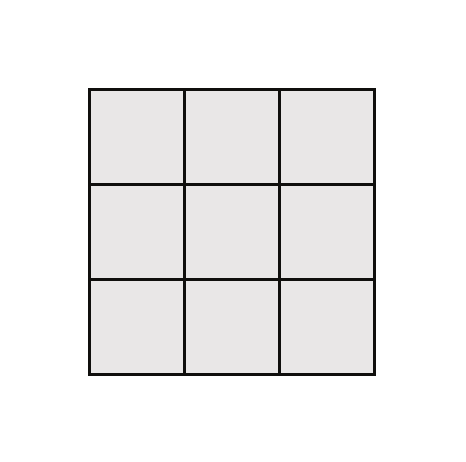}
		\label{3B3}}\hspace{-9mm}
	\subfigure[3 by 4]{
		\includegraphics[width=1.27in]{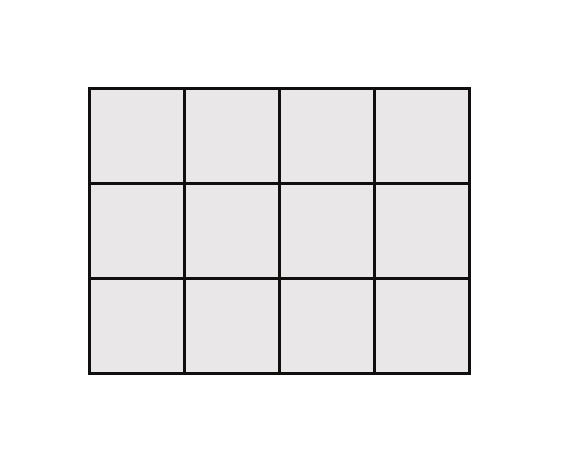}
		\label{3B4}}
	\caption{The $N\ \mathrm{by}\ M$ rooms layouts.}
	\label{NBM}
\end{figure}

\begin{figure}[t!]
	\centering
	\subfigure[CDFs of $\tau$ in a $3\ \mathrm{by}\ 2$ scenario.]{
		\includegraphics[width=3in]{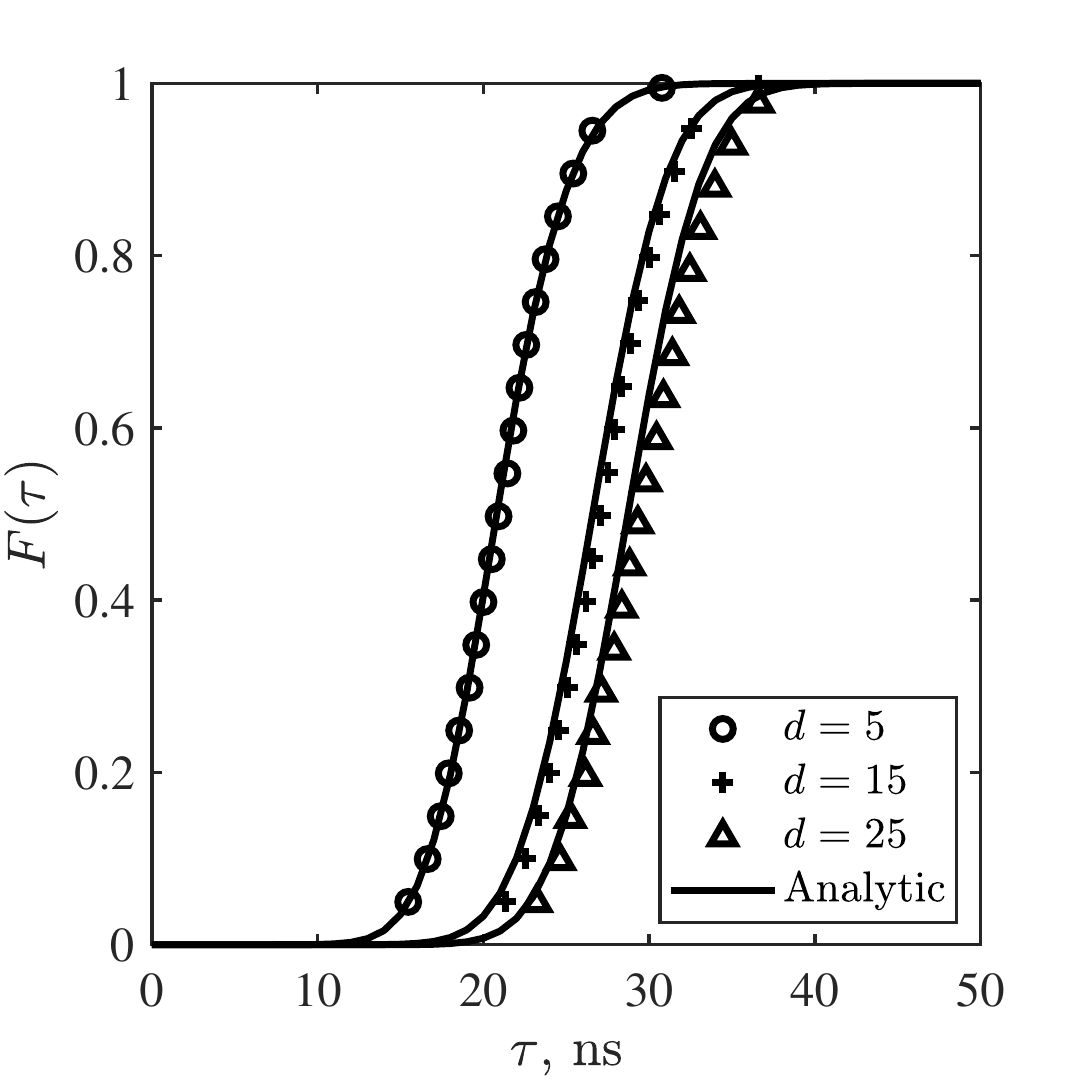}\label{taucdf_3B2}}\\
	\subfigure[CDFs of $\tau$ in a $3\ \mathrm{by}\ 3$ scenario.]{
		\includegraphics[width=3in]{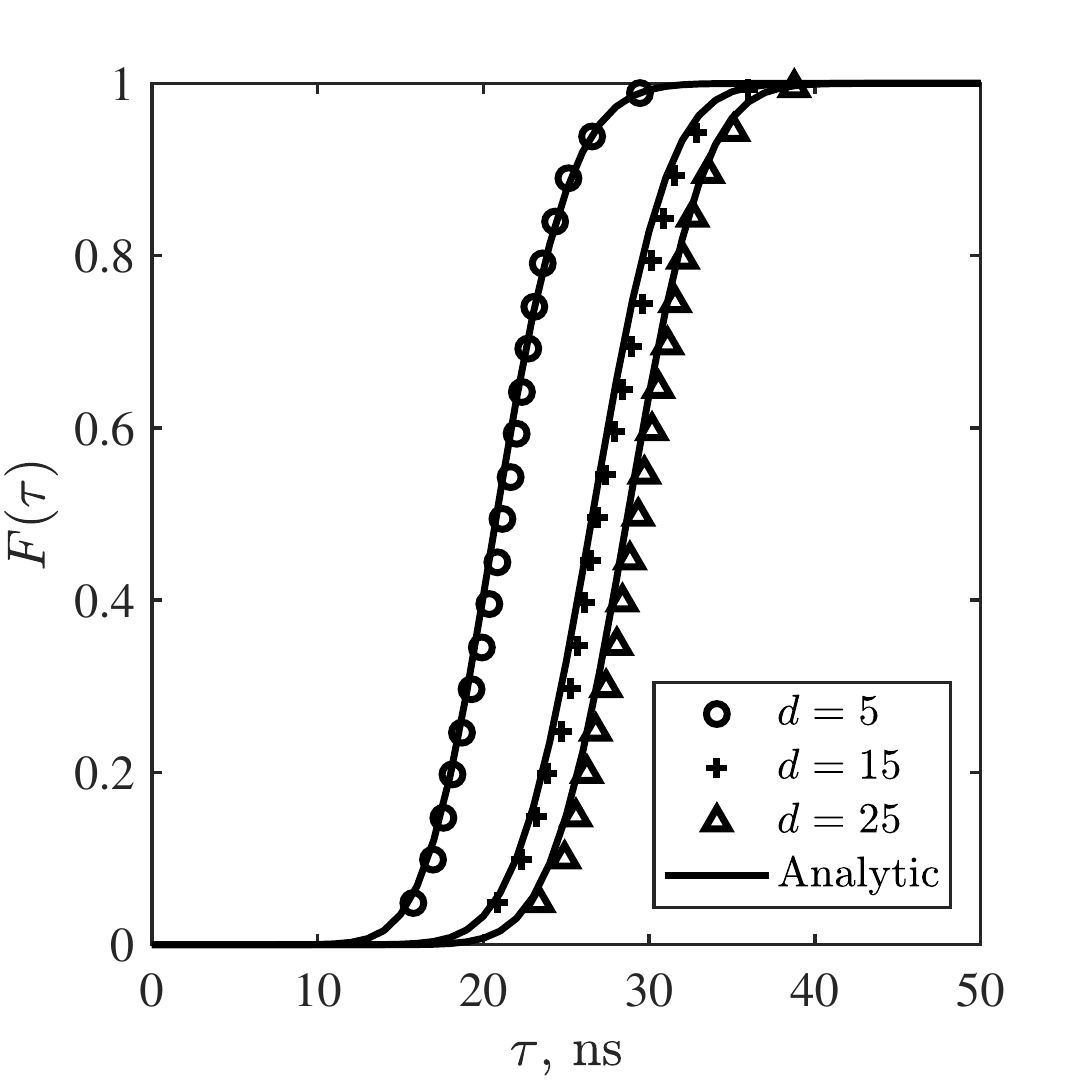}
		\label{taucdf_3B3}}
	\caption{CDF of $\tau$ with different $d$ in $3\ \mathrm{by}\ 2$ and $3\ \mathrm{by}\ 3$ scenarios.}
	\label{taucdf}
\end{figure}

\begin{figure}[t!]
	\centering
	\subfigure[PDF of $d$ in a $3\ \mathrm{by}\ 2$ scenario.]{
		\includegraphics[width=3in]{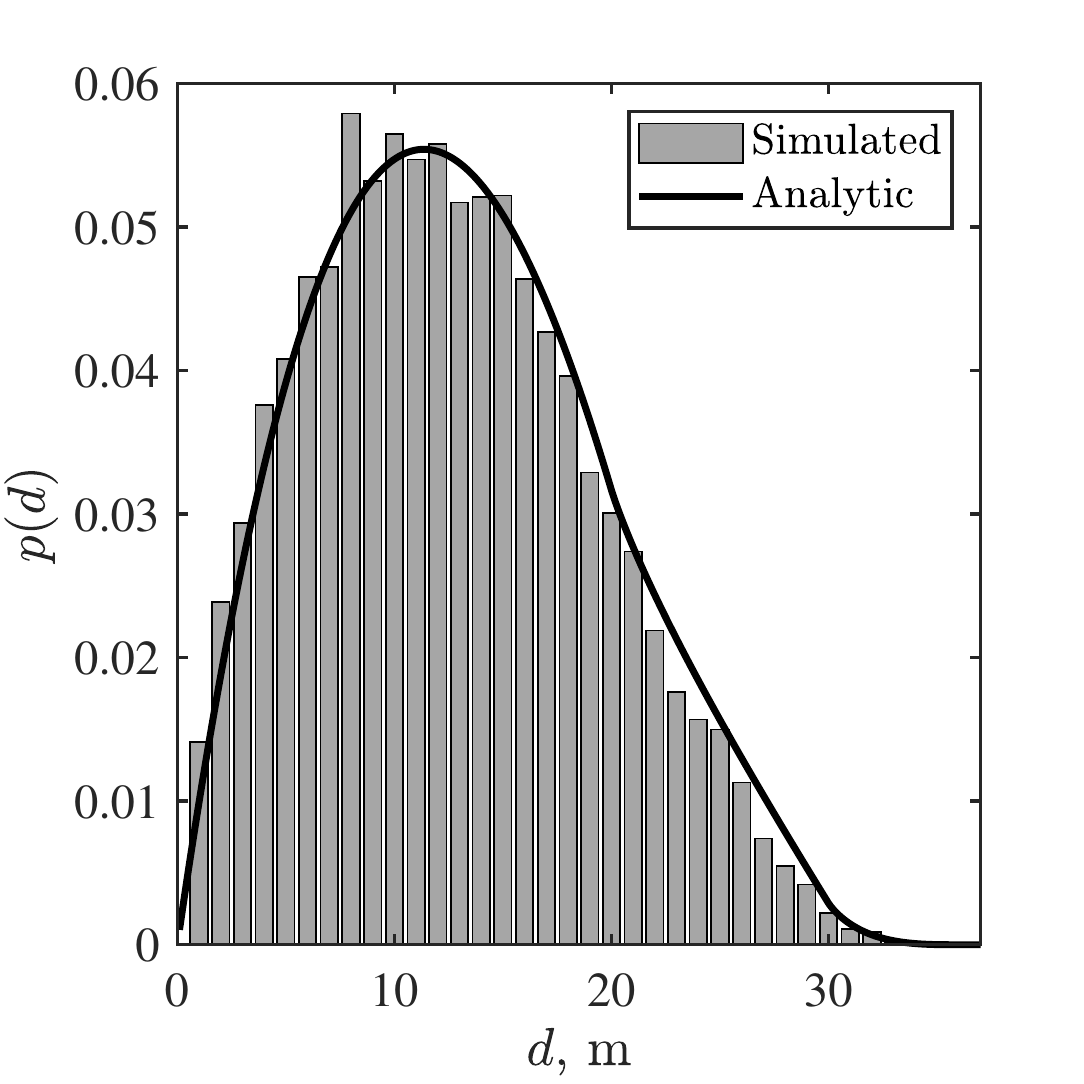}\label{distance_distribution_3B2}}\\
	\subfigure[PDF of $d$ in a $3\ \mathrm{by}\ 3$ scenario.]{
		\includegraphics[width=3in]{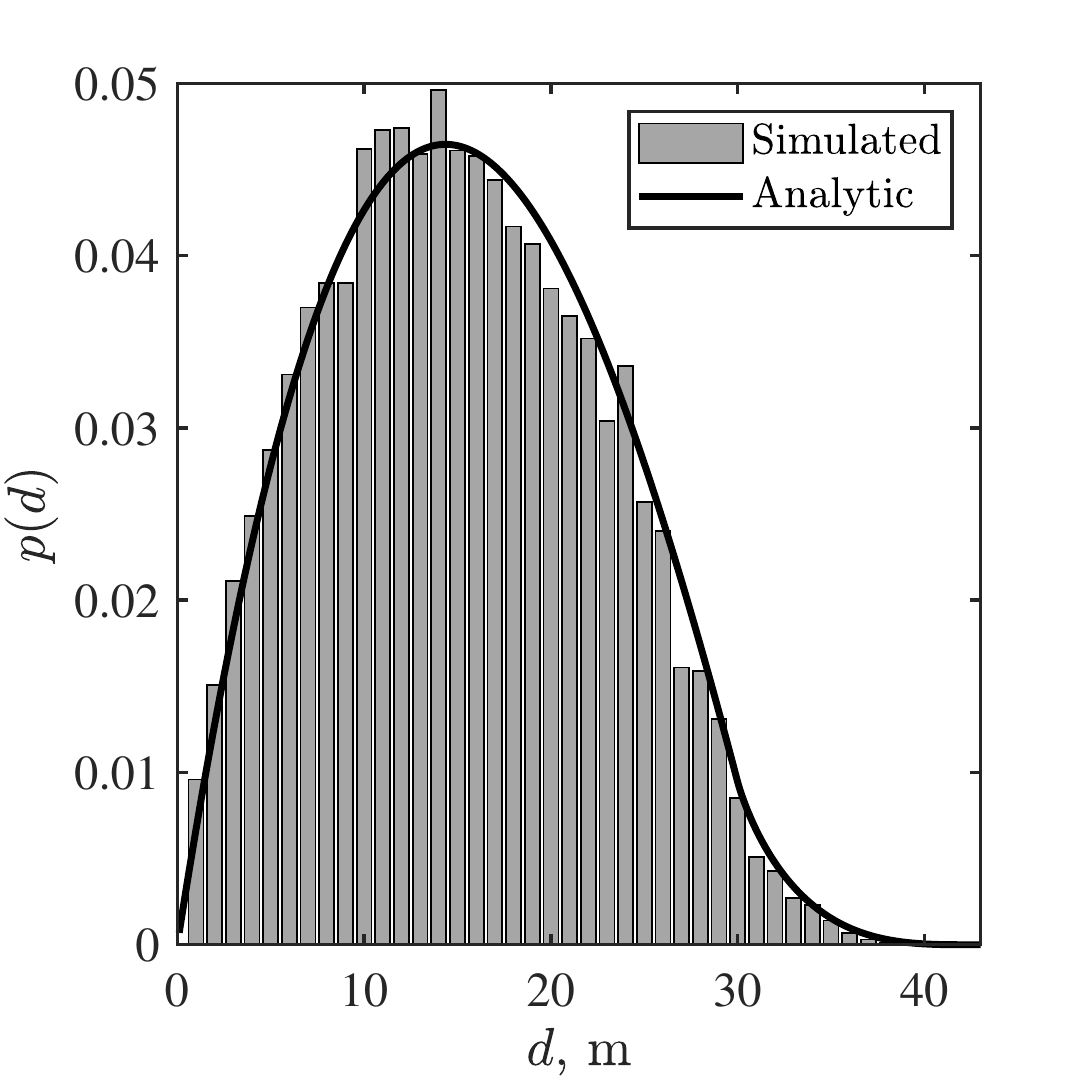}
		\label{distance_distribution_3B3}}
	\caption{PDF of $d$ in $3\ \mathrm{by}\ 2$ and $3\ \mathrm{by}\ 3$ scenarios.}
	\label{discdf}
\end{figure}

\begin{figure}[t!]
	\centering
	\subfigure[{\color{black}Comparisons in a $3\ \mathrm{by}\ 2$ scenario.}]{
		\includegraphics[width=3in]{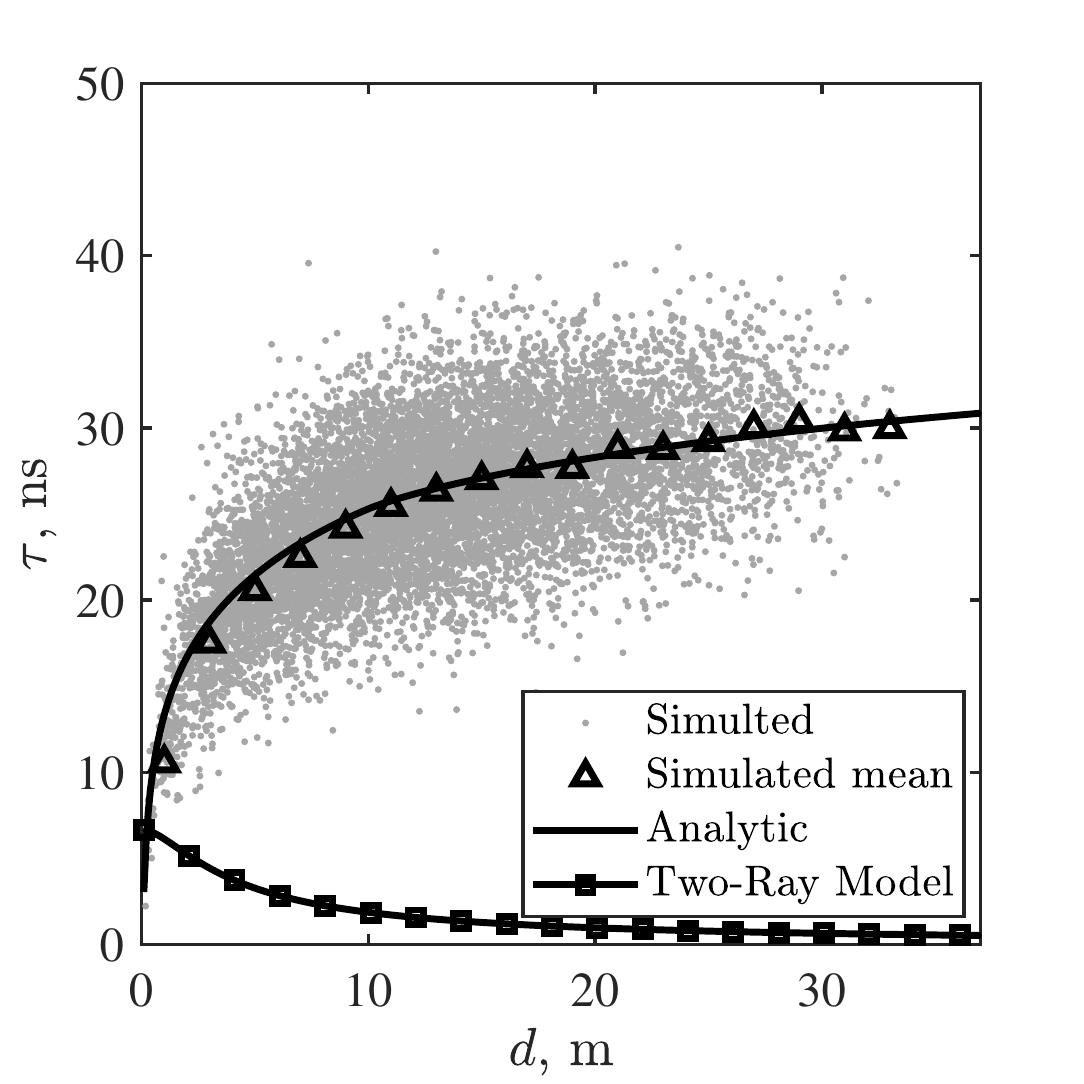}\label{rms_3B2}}\\
	\subfigure[Comparisons in a $3\ \mathrm{by}\ 3$ scenario.]{
		\includegraphics[width=3in]{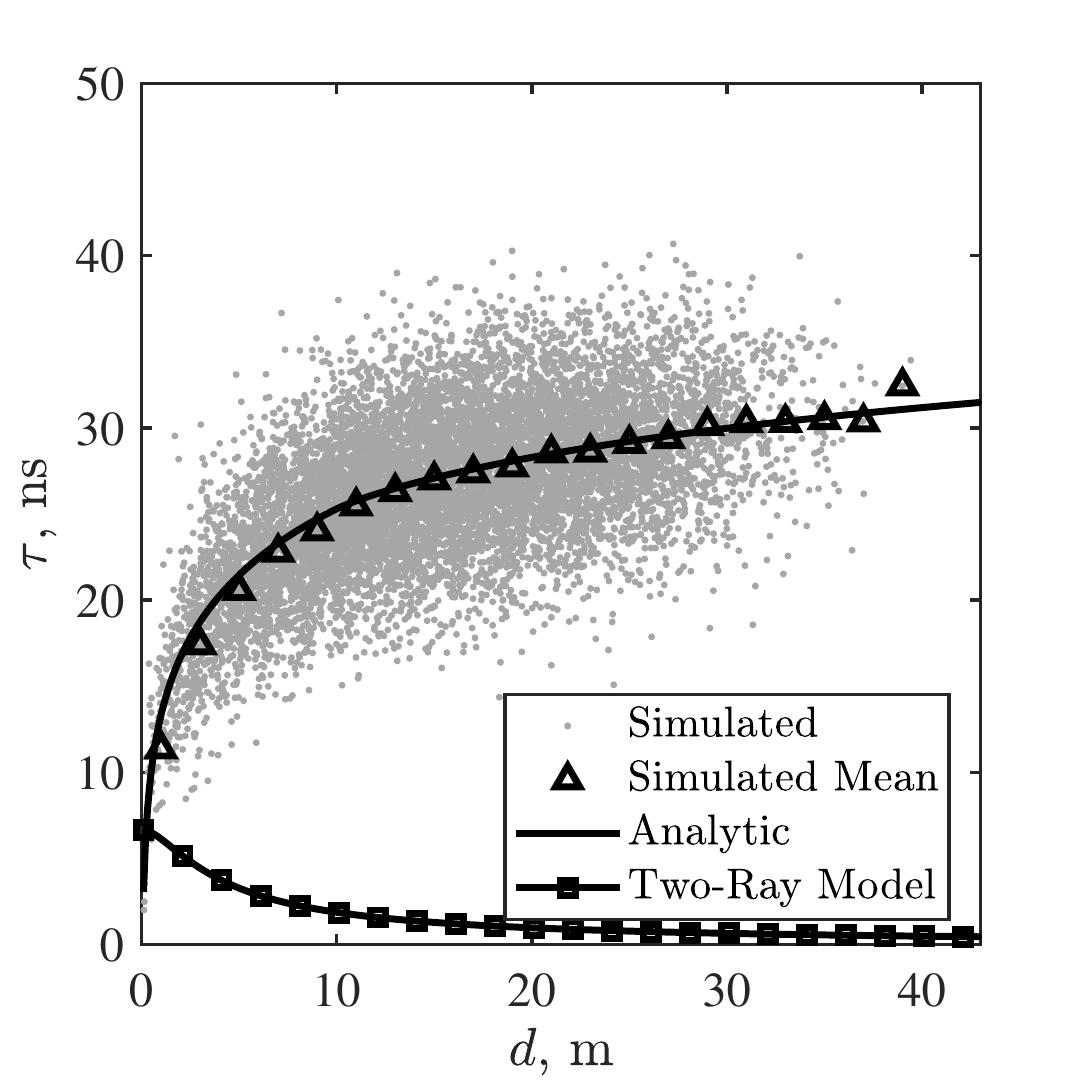}
		\label{rms_3B3}}
	\caption{The simulated values of {\color{black}$\tau_\mathrm{I}$} are compared with the analytic value and the {\color{black}$\tau_\mathrm{O}$} values for corresponding two-ray models.}
	\label{rms_NBM}
\end{figure}

\begin{figure}[t!]
	\centering
	\subfigure[Comparison in scenarios with different room areas.]{
		\includegraphics[width=3in]{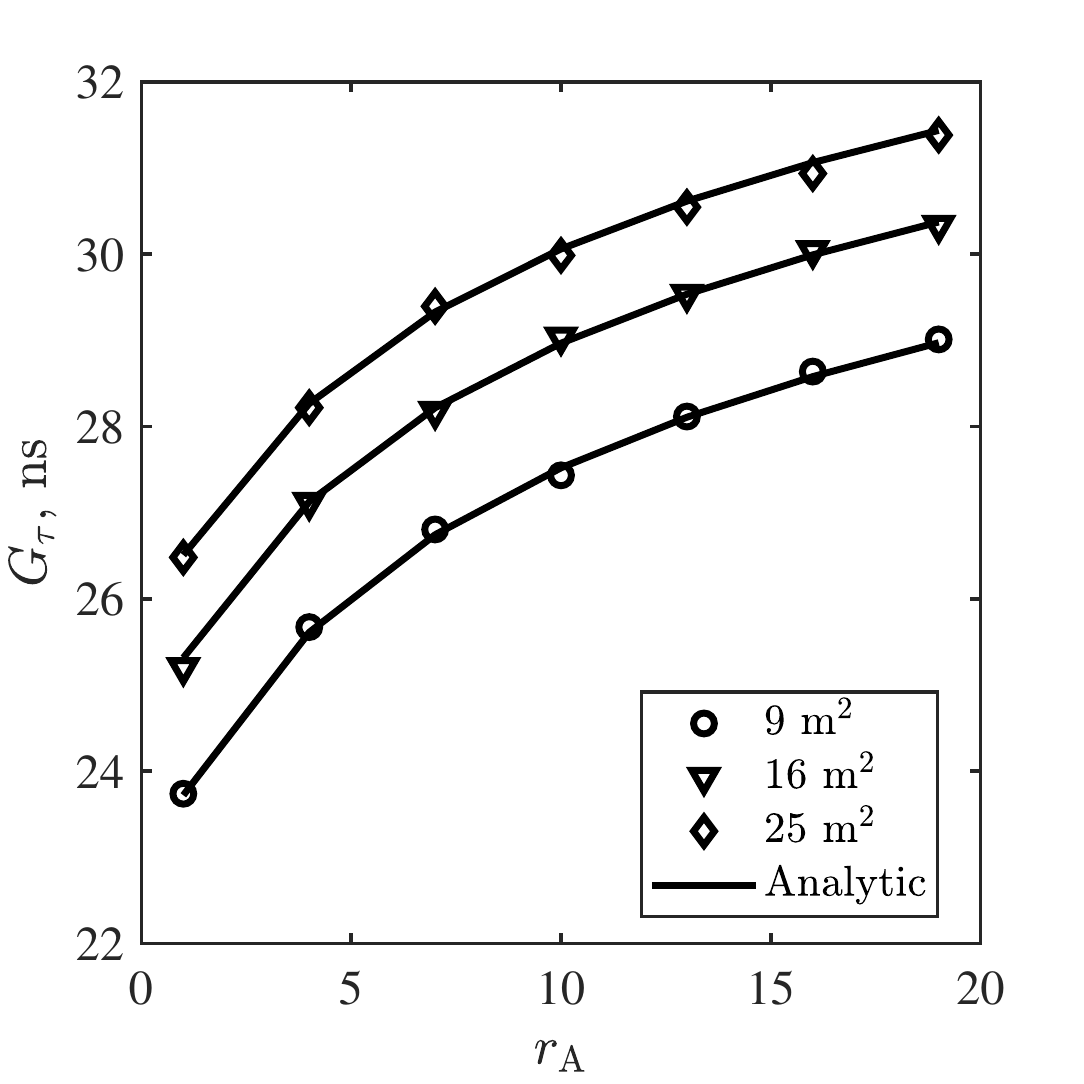}\label{difardifs}}\\
	\subfigure[Comparison in scenarios with different room layouts.]{
		\includegraphics[width=3in]{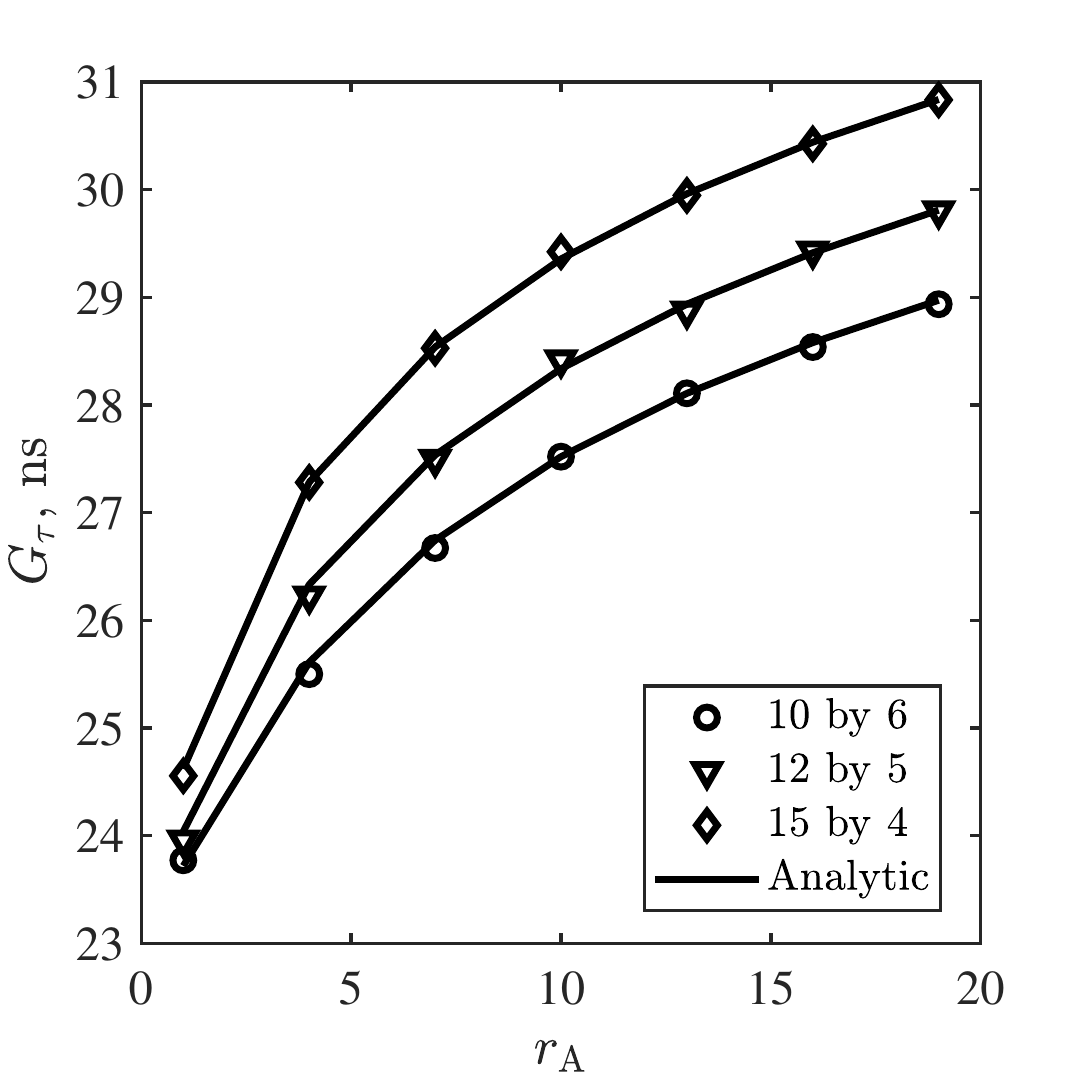}
		\label{difardifroom}}
	\caption{Comparison of $G_\tau$ with different aspect ratio $r_\mathrm{A}$ in different scenarios.}
	\label{difar}
\end{figure}

\begin{figure}[t!]
	\centering
	\subfigure[Comparison in scenarios with different room areas.]{
		\includegraphics[width=3in]{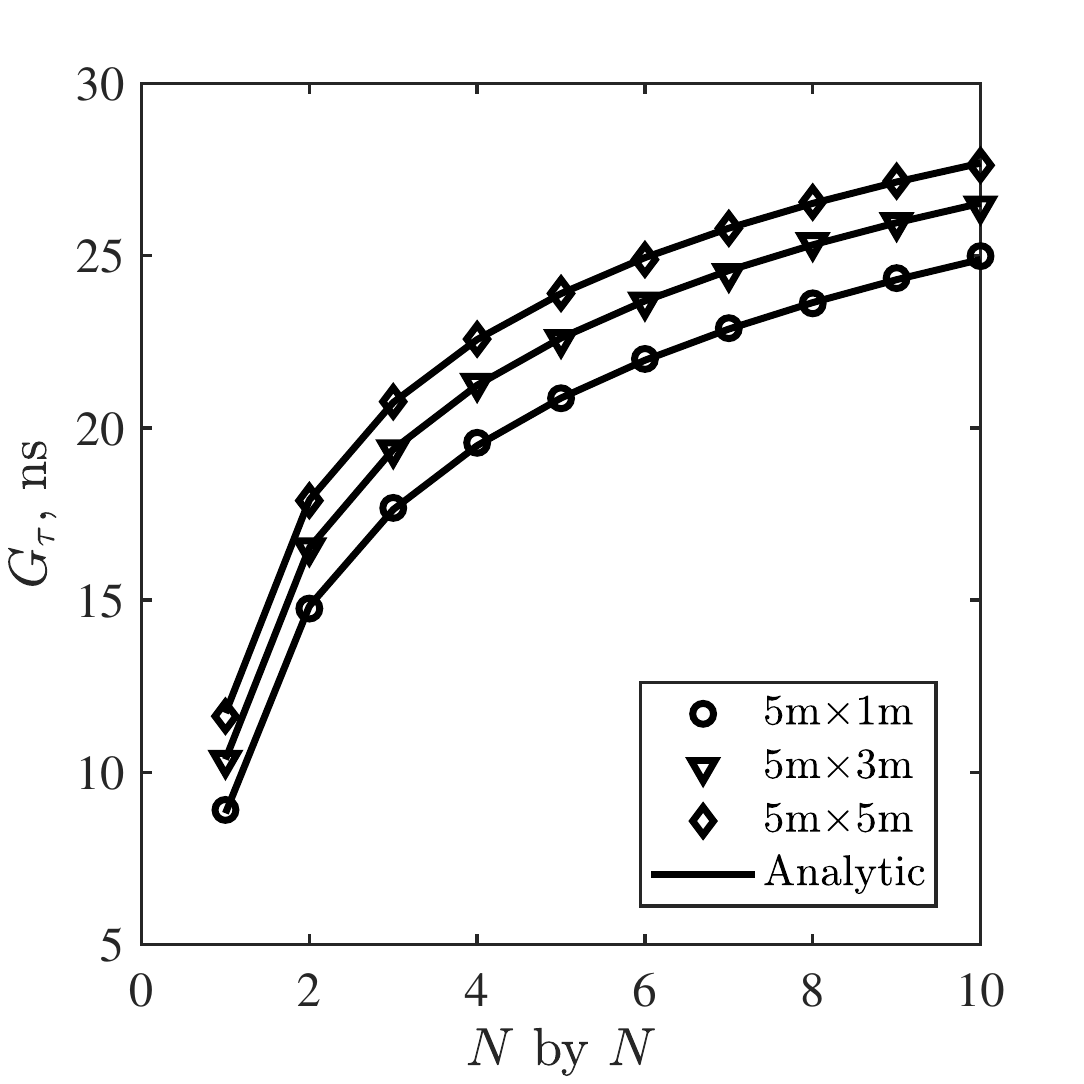}\label{difroomareadifdia}}\\
	\subfigure[Comparison in scenarios with same diagonal but different aspect ratio.]{
		\includegraphics[width=3in]{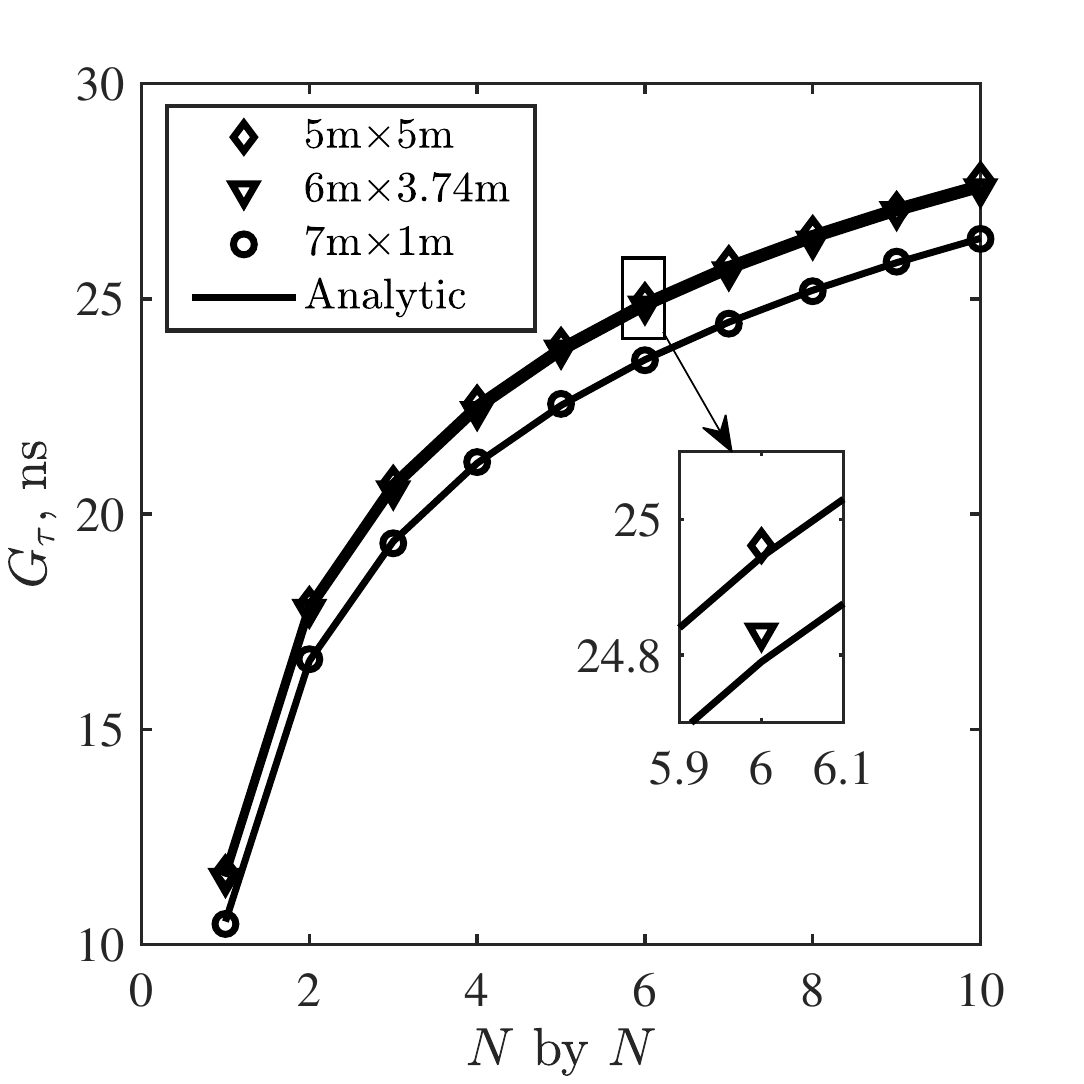}
		\label{difroomareasamedia}}
	\caption{Comparison of $G_\tau$ with different room layouts in different scenarios.}
	\label{difroomarea}
\end{figure}

\begin{figure}[t!]
	\centering
	\subfigure[Comparison in a floor with $N\ \mathrm{by}\ N$ room layouts.]{
		\includegraphics[width=3in]{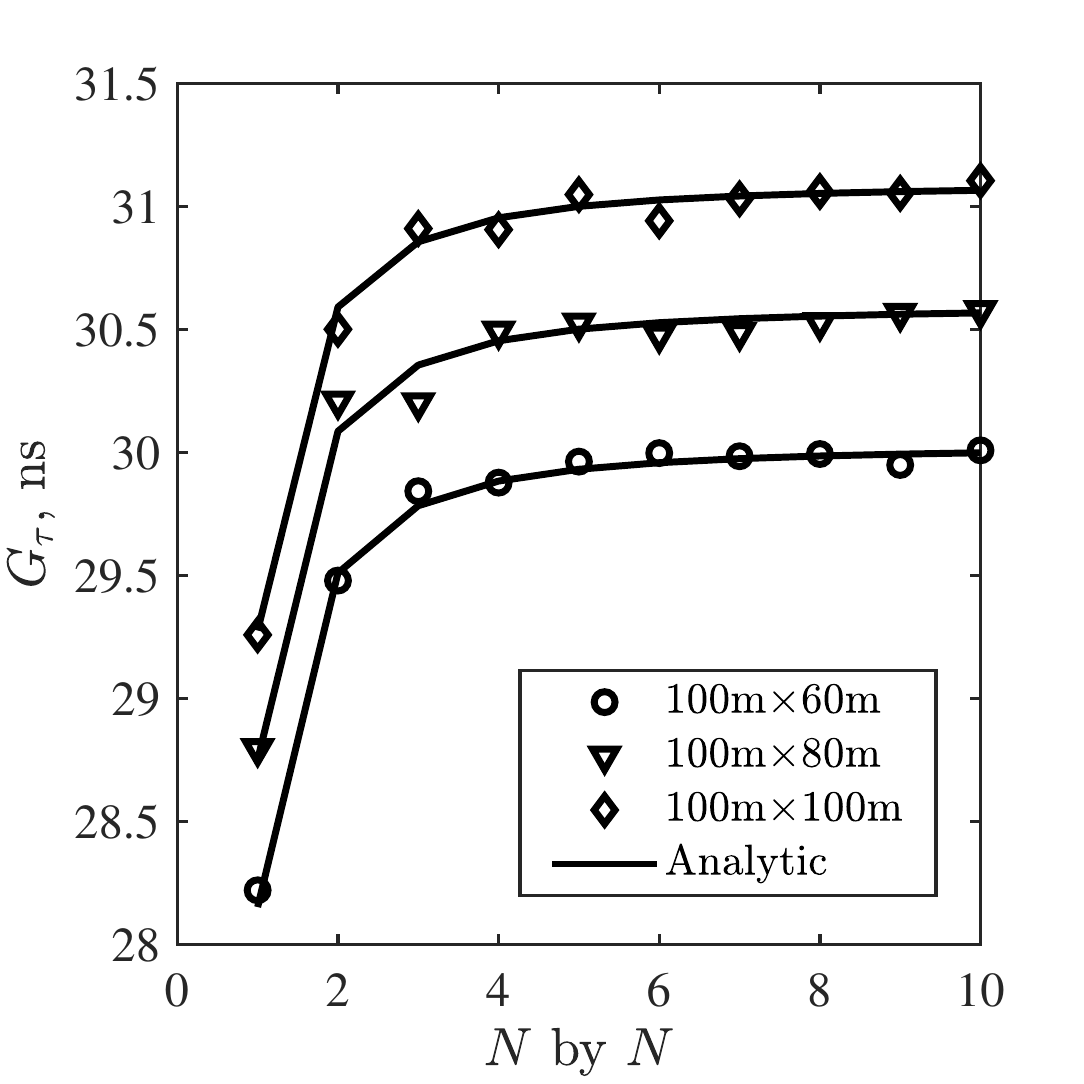}\label{samefloorNBN}}\\
	\subfigure[Comparison in a floor with $N\ \mathrm{by}\ 2N$ room layouts.]{
		\includegraphics[width=3in]{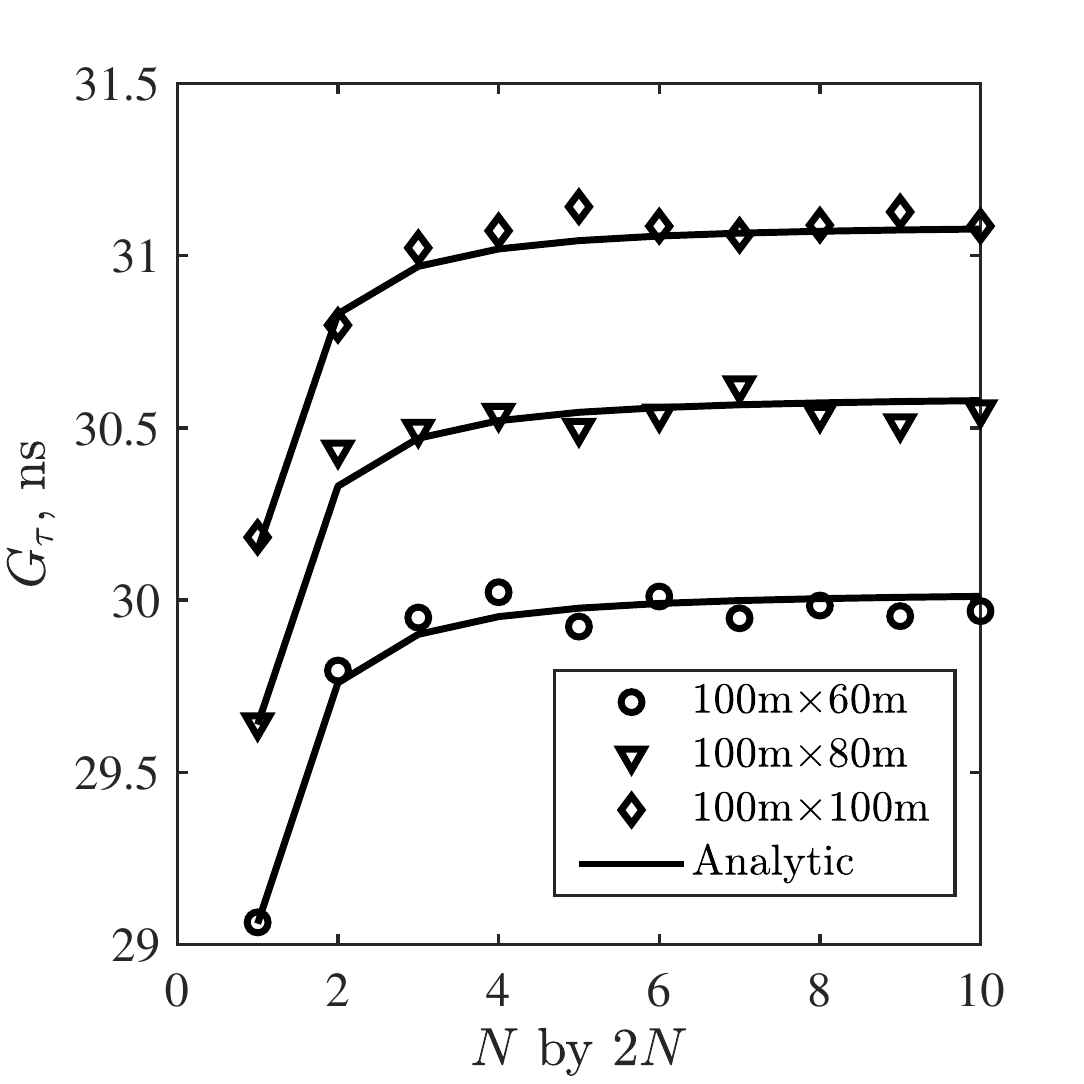}
		\label{samefloorNB2N}}
	\caption{Comparison of $G_\tau$ with different room layouts within the same-sized floor}
	\label{samefloor}
\end{figure}

\begin{figure}[t!]
	\centering
	\subfigure[Comparison in a floor with $N\ \mathrm{by}\ N$ room layouts.]{
		\includegraphics[width=3in]{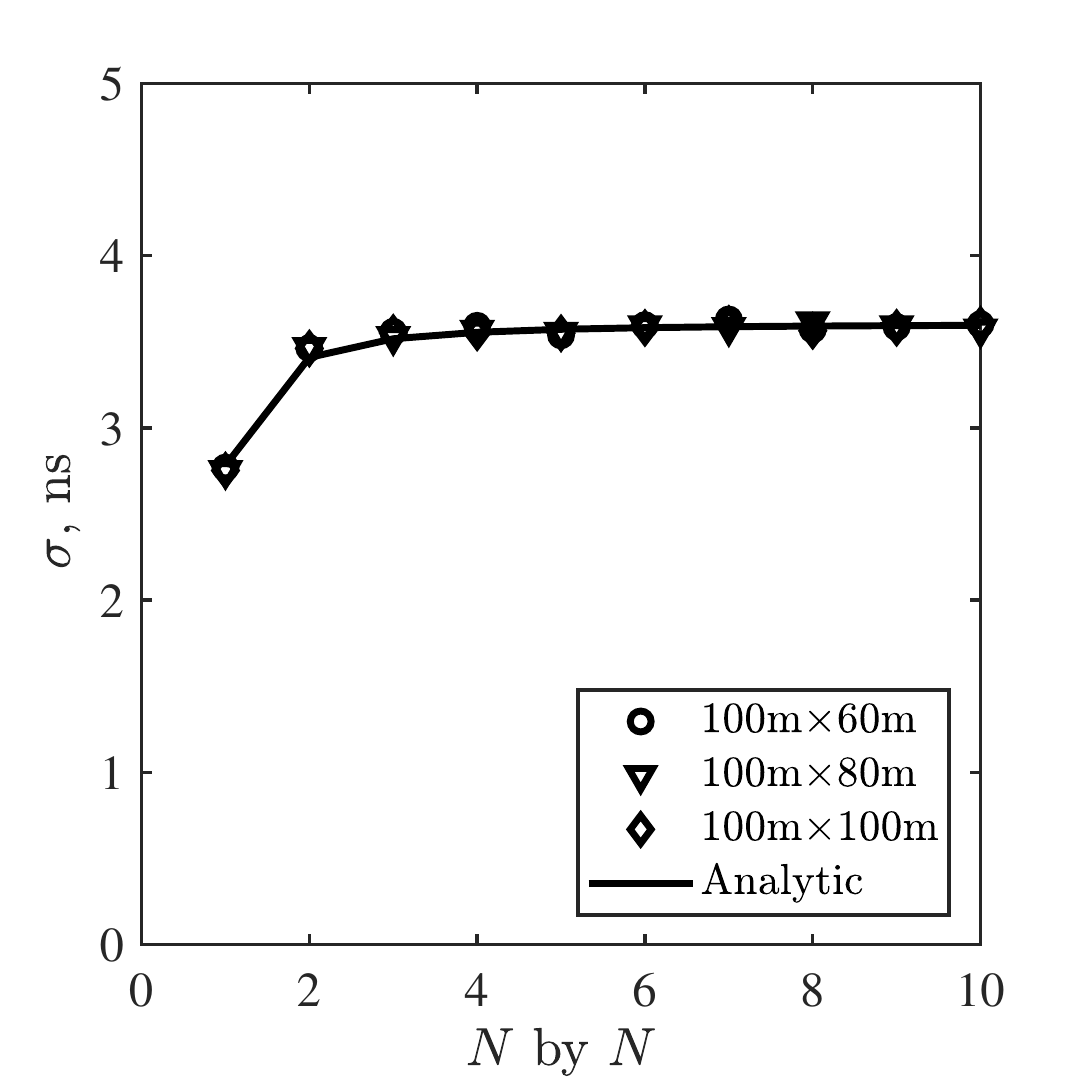}\label{samefloorNBN_sigma}}\\
	\subfigure[Comparison in a floor with $N\ \mathrm{by}\ 2N$ room layouts.]{
		\includegraphics[width=3in]{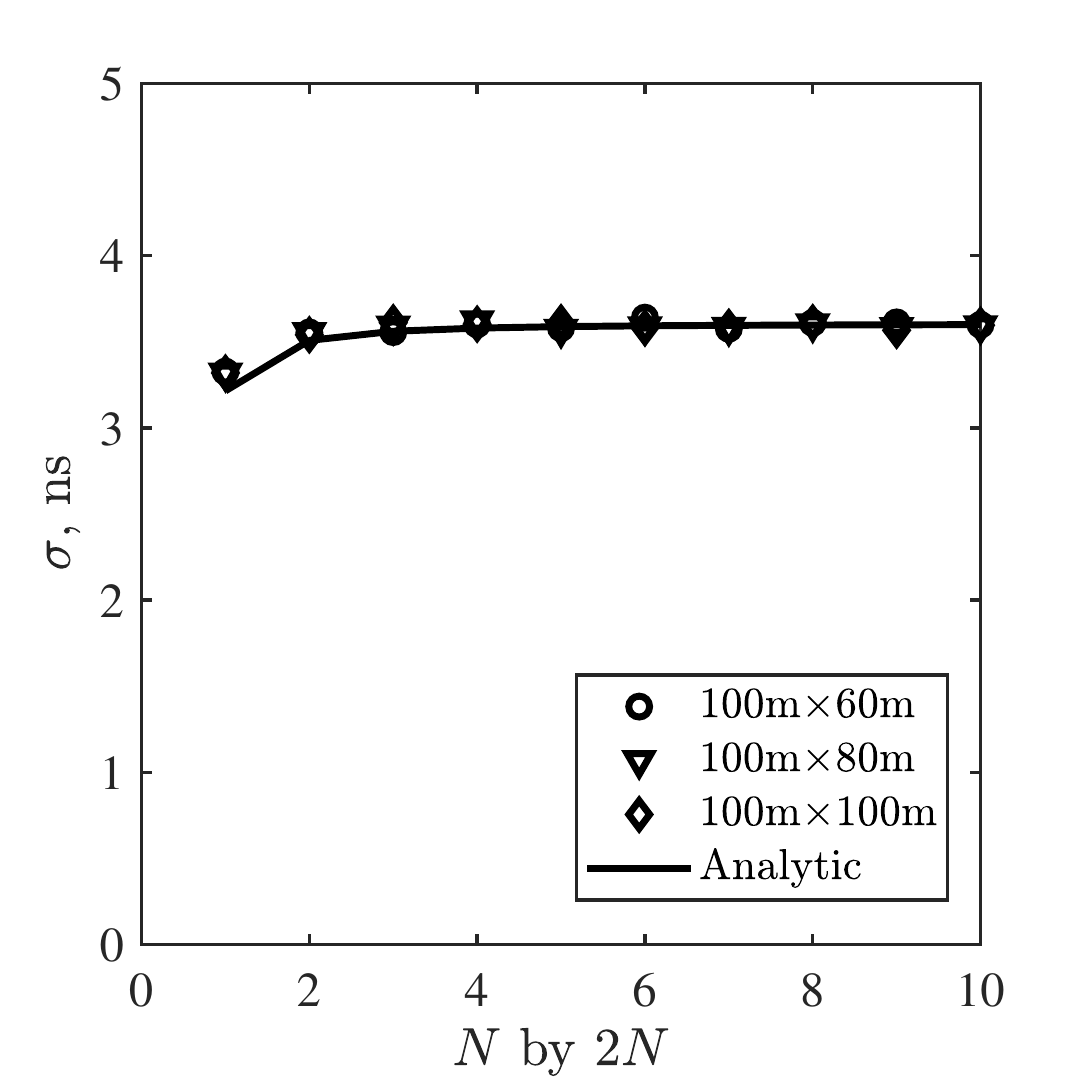}
		\label{samefloorNB2N_sigma}}
		\caption{{\color{black}Comparison of $\sigma$ with different room layouts within the same-sized floor}}
	\label{samefloor_sig}
\end{figure}

\begin{figure}[!t]
	\centering
	\includegraphics [width=3in]{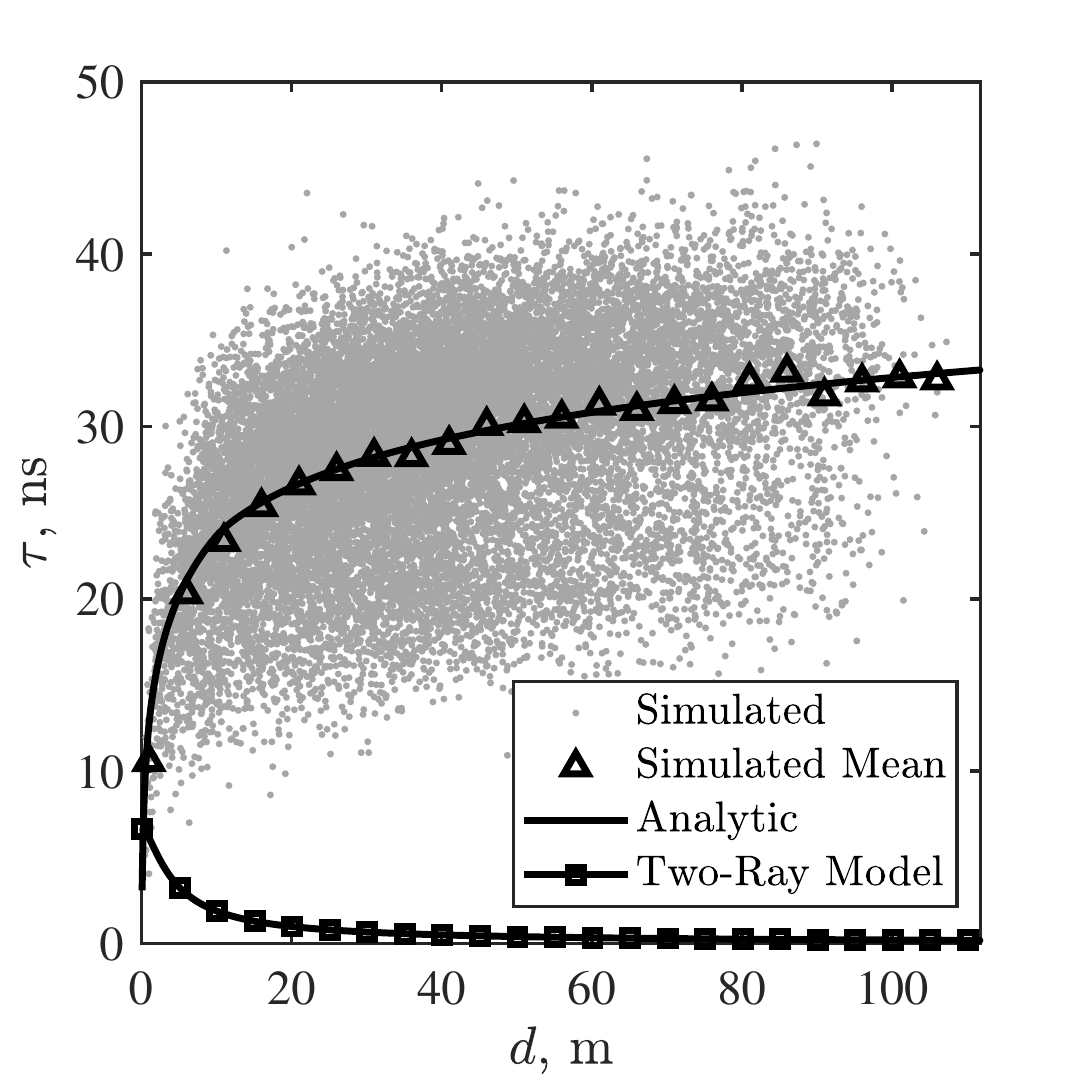}
	\caption{The simulated values of {\color{black}$\tau_\mathrm{I}$} are compared with the analytic value of the Winner II A1 scenario and the {\color{black}$\tau_\mathrm{O}$} values for corresponding two-ray models.}
	\label{Winner2compare}
	\centering
\end{figure}

%

\section{Numerical Results}

For validating the model's reliability, the Monte-Carlo simulations are first performed in room layouts of $3\ \mathrm{by}\ 2$ and $3\ \mathrm{by}\ 3$. Fig. \ref{NBM} illustrates the naming conventions for a single type of room layout, e.g., $N\ \mathrm{by}\ M$ denotes a room layout consisting of $N$ rows and $M$ columns. $h_\mathrm{t}$ is set as $4\ \mathrm{m}$ and $h_\mathrm{r}$ is set as $3\ \mathrm{m}$ for all scenarios. The model is then verified step by step.
In every Monte-Carlo simulation, we randomly pick 10000 positions in the scene to position transmitters and receivers, respectively, and then established one-to-one communication links for them. 

The CDFs of RMS-DS for $3\ \mathrm{by}\ 2$ and $3\ \mathrm{by}\ 3$ room layouts at various transceiver distances ($d=5\ \mathrm{m}$, $15\ \mathrm{m}$, and $25\ \mathrm{m}$) are compared to the analytical values determined by \eqref{eq15} in Fig. \ref{taucdf}.
As can be observed, the statistical results closely correspond to the simulated values in different layouts.
The images demonstrate that the expectation value of $\tau_\mathrm{I}$ grows as the transceiver distance $d$ increases and that the increase is more pronounced with a shorter $d$. 

Fig. \ref{discdf} illustrates the PDFs of $d$ in $3\ \mathrm{by}\ 2$ and $3\ \mathrm{by}\ 3$ room layouts when the transmitter and receiver positions are randomly chosen.
The statistical values are accurate when compared to the simulated results.

The $\tau_\mathrm{I}$ values for all links simulated using the Monte-Carlo method are shown in Fig. \ref{rms_NBM} in comparison to the analytic solution of $f_{\tau_\mathrm{I}}(d)$. Fig. \ref{rms_NBM} also shows the expectation value of simulation results in $\tau$-dimension with the interval of $2$m, represented by markers `$\bigtriangleup$'.
Additionally, in Fig. \ref{rms_NBM} we display the $f_{\tau_\mathrm{I}}(d)$ values for these distances in an open environment.
This picture confirms the accuracy of the proposed model and cross-validates the relationship between $f_{\tau_\mathrm{I}}(d)$ and distance $d$ combined with Fig. \ref{taucdf}.
At shorter $d$, $f_{\tau_\mathrm{I}}(d)$ increases with $d$ faster.
This may be because the increase in $d$ at this stage dramatically increases the complexity of the communication environment, i.e., a small increase in distance may cause the link of transition from LOS to NLOS.
Furthermore, when $d$ is excessively long, $f_{\tau_\mathrm{I}}(d)$ increases slowly with $d$.
This is because the communication environment is already harsh: increasing the distance between nodes will merely result in PL growth and will have little effect on the LOS/NLOS conditions.
The value of $\tau_\mathrm{I}$ in an ideal open area diminishes with increasing $d$ because the distance gap between direct and reflected routes becomes less apparent.


Following that, we placed the single-type rooms in a number of layouts to demonstrate the model's universality and to make a preliminary summary of the effect of room layout on channel DS performance.
The Figs. \ref{difardifs} and \ref{difardifroom} illustrate the effect of room aspect ratios to $G_\tau$ when the unit rooms areas are fixed. 
In Fig. \ref{difardifs}, the same layouts of $10\ \mathrm{by}\ 6$ rooms are employed, and the $G_\tau$ of identical aspect ratio $r_\mathrm{A}$ rooms with varying room sizes are also compared longitudinally.
As shown, when the unit room area is fixed, $G_\tau$ increases as $r_\mathrm{A}$ grows.
Additionally, when the $r_\mathrm{A}$ is fixed, the larger unit room area results in a larger $G_\tau$.

In Fig. \ref{difardifroom}, $60$ same $9\ \mathrm{m}^2$ rooms is employed, and the $G_\tau$ of identical aspect ratio $r_\mathrm{A}$ rooms with varying room layouts ($15\ \mathrm{by}\ 4$, $12\ \mathrm{by}\ 5$ and $10\ \mathrm{by}\ 6$) are also compared longitudinally.
As shown, when both unit area and total area remain unchanged, room layouts will also affect $G_\tau$, i.e., the floor with a bigger aspect ratio can cause a greater $G_\tau$.
What these two phenomena have in common is that a higher aspect ratio with the same room area or a larger room area with the same room layout can lead to a longer communication link within the building.
Because the longer the distance, the bigger the RMS-DS, and hence the greater the RMS-DS expectation of the building in these instances.
Fig.\ref{difardifroom} also demonstrates that as rooms are placed to make the building longer and narrower, the $G_\tau$ increases.
This is also because a narrower building with the same area has a higher probability of having a longer transmission distance. 

Similar influences apply to Fig. \ref{difroomareadifdia}.
Fig. \ref{difroomareadifdia} illustrates the changes in $G_\tau$ induced by similar room arrangements with the varying number of rooms and vertically compares $G_\tau$ with the same room layouts but different unit room areas.
Similarly, with a constant room area, adding more rooms increases $G_\tau$ substantially.
On the one hand, the increase in room number, along with the increase in building area, results in an increase in the possible transmission distance. On the other hand, more rooms result in a greater probability of NLOS transmissions.
Vertically comparing, increasing the room areas increases $G_\tau$ in this figure, which is also related to the increased maximum transceiver distance. 

To shield the effect of varying maximum propagation lengths on $G_\tau$, we fixed the diagonal lengths of the unit rooms and compared $G_\tau$ again, as shown in Fig. \ref{difroomareasamedia}. It shows that under the condition that the diagonal length, namely the maximum communication distance, is fixed, $G_\tau$ will still be affected by the building area --- the smaller the building area, the better the $G_\tau$ performance.

Finally, we partitioned a fixed structure into $N\ \mathrm{by}\ N$ and $N\ \mathrm{by}\ 2N$ equal sections and compared their $G_\tau$, as seen in Fig. \ref{samefloorNBN} and Fig. \ref{samefloorNB2N} respectively.
When a building follows a set contour, adding rooms results in a larger $G_\tau$.
This effect is more noticeable when there are fewer rooms and less noticeable when there are enough rooms because the majority of the building's links have been NLOS at this stage, and thus adding rooms has little influence on RMS-DS. 

Combined with the above analysis, we discover that the diagonal length and the area of the building are the primary determinants of the assessed building's $G_\tau$. When a building has a longer diagonal length or a larger area, it will have a greater $G_\tau$.
When there are fewer rooms, the increase of walls substantially raises $G_\tau$, but this element becomes insignificant when there are more rooms.

{\color{black}
To observe the correlation between the model's reliability and the BUD's inner structure, we compared the simulation results and statistical results of $\sigma$ of different room layouts in a same-sized building, as shown in Fig. \ref{samefloor_sig}. Fig. \ref{samefloor_sig} illustrates that similar to the change in $G_\tau$, $\sigma$ becomes larger as the number of rooms $N_\mathrm{r}$ increases. However, the $\sigma$ values were all between $2.7$ and $3.6$, and the increase in $\sigma$ was not significant as $N_\mathrm{r}$ becomes large. This is synchronized with the $\sigma_{\mathrm{LOS},\upsilon_i,\tau_\mathrm{I}}$ and $\sigma_{\mathrm{NLOS},\upsilon_i,\tau_\mathrm{I}}$ computed according to \eqref{RMS:sigma}. When there are fewer rooms in the BUD, $\sigma$ is more inclined to $\sigma_{\mathrm{LOS},\upsilon_i,\tau_\mathrm{I}}$. Conversely, when there are more rooms, $\sigma$ is close to $\sigma_{\mathrm{NLOS},\upsilon_i,\tau_\mathrm{I}}$. This demonstrates that the uncertainty of the indoor RMS-DS mainly determines the deviation in the proposed evaluation model. However, the statistical analysis of the RMS-DS for a whole BUD is not affected by the deviation of an individual link, and therefore the $G_\tau$ as shown in Figs. \ref{difar}-\ref{samefloor} does not deviate significantly from the simulated results.
}

Finally, we validated the model by comparing the simulated and analytical values of $G_\tau$ for a real indoor scenario. This scenario is a $50\ \mathrm{m}\times 100\ \mathrm{m}$ indoor scenario referred from Winner II A1 scenario \cite{meinila2009winner}, which consists of 40 identical $10\ \mathrm{m}\times 10\ \mathrm{m}$ offices and two $5\ \mathrm{m}\times 100\ \mathrm{m}$ corridors.
The simulated result is $G_{\tau,\mathrm{sim}}=27.7866 \ \mathrm{ns}$, and the analytical result is $G_{\tau,\mathrm{ana}}=27.7435\ \mathrm{ns}$. Detailed simulation results are shown in Fig. \ref{Winner2compare}. These results demonstrate that the proposed model can also accurately and quickly evaluate the actual complex building structures.

\section{Conclusion}

In this paper, we have defined the metric of DS gain to build the BWP evaluation scheme of channel DS.
The DS gain, $G_\tau$, is the metric defined by comparing the expected RMS-DS within a building to the RMS-DS in an open space with the same topography.
We first derived the analytical probability distribution model of the RMS-DS of the entire building with the transceiver distance by integrating the RMS-DS model of a single room with the room arrangement of a building.
Then, we derived the PDF of the transceiver distance within the building and the analytical solution for calculating $G_\tau$. {\color{black}A metric $\sigma$ was also defined and derived to assess model reliability.}
Finally, we obtained $G_\tau$ values in various room layouts through simulation and compared them to those of the analytical model, and demonstrated that the proposed model could accurately evaluate the channel DS performance of a building under design.
Our work provides the channel DS evaluation guidelines for future building design, which is an important aspect of the BWP evaluation framework. As the RMS-DS has a profound impact on wireless system design, the work is very relevant to future generations of wireless systems that need to support services with extreme requirements on latency, reliability and data rates.

{\color{black}In the future, we will also consider the impact of the outdoor BSs on the building RMS-DS performance.	This will provide architects with suggestions for building design in another aspect.
Additionally, how the complexity and effectiveness of the pilot design are affected by in-building wireless communication environments with different $G_\tau$ will be another future research direction.}





\ifCLASSOPTIONcaptionsoff
  \newpage
\fi

\bibliography{IEEEfull.bib}
\bibliographystyle{IEEEtran}  

\end{document}